\documentclass[
preprint,
aps,prb,
superscriptaddress,
showpacs,
longbibliography
]{revtex4-1}
\usepackage{graphicx}
\usepackage{amssymb,amsthm,amsmath}
\usepackage{epstopdf}
\usepackage{pdfsync}
\usepackage{graphicx}
\usepackage{placeins}
\usepackage{braket}
\usepackage{physics}

\def\urlprefix{}

\usepackage[position=t]{subcaption}
\usepackage{natbib}

\newcommand{\beq}{\begin{equation}}
\newcommand{\eeq}{\end{equation}}
\newcommand{\beqn}{\begin{equation*}}
\newcommand{\eeqn}{\end{equation*}}

\newcommand{\ybyvo}{$^{171}$Yb$^{3+}$:YVO$_4$~}

\newcommand{\ybyso}{$^{171}$Yb$^{3+}$:$\mathrm{Y_2SiO_5}$~}
\newcommand{\yvo}{YVO$_4$}
\newcommand{\yso}{$\mathrm{Y_2SiO_5}$~}
\newcommand{\ybion}{ ${}^{171}\mathrm{Yb}^{3+}$~}
\newcommand{\eryso}{$^{167}$Er$^{3+}$:$\mathrm{Y_2SiO_5}$~} 

\newcommand{\nucspintrans}{ $\ket{1'}_g \rightarrow \ket{2'}_g $}
\newcommand{\opttransvo}{ $\ket{2'}_g \rightarrow \ket{1'}_e $}
\newcommand{\opttransvr}{ $\ket{1'}_g \rightarrow \ket{1'}_e $}

\newcommand{\eq}[1]{Eq.~(\ref{eq:#1})}
\newcommand{\fig}[1]{Fig.~\ref{fig:#1}}

\newcommand{\sect}[1]{Section~\ref{sec:#1}}

\newcommand{\kavli}{Kavli Nanoscience Institute and Thomas J. Watson, Sr., Laboratory of Applied Physics, California Institute of Technology, Pasadena, California 91125, USA}
\newcommand{\iqim}{Institute for Quantum Information and Matter,
California Institute of Technology, Pasadena, California 91125, USA}
\newcommand{\bozeman}{Department of Physics, Montana State University, Bozeman, Montana 59717, USA}
\newcommand{\chicago}{Institute of Molecular Engineering, University of Chicago, Chicago, IL 60637 USA}
\bibpunct{[}{]}{;}{n}{}{}

\begin{document}

\title{Characterization of $\mathbf{{}^{171}Yb^{3+}\!:\! YVO_4}$ for photonic quantum technologies}
\author{Jonathan M. Kindem}
\affiliation{\kavli}
\affiliation{\iqim}
\author{John G. Bartholomew}
\affiliation{\kavli}
\affiliation{\iqim}
\author{Philip J. T. Woodburn}
\affiliation{\bozeman}
\author{Tian Zhong}
\affiliation{\kavli}
\affiliation{\iqim}
\affiliation{\chicago}
\author{Ioana Craiciu}
\affiliation{\kavli}
\affiliation{\iqim}
\author{Rufus L. Cone}
\affiliation{\bozeman}
\author{Charles W. Thiel}
\affiliation{\bozeman}
\author{Andrei Faraon}
\email[Email address: ]{faraon@caltech.edu}
\affiliation{\kavli}
\affiliation{\iqim}
\date{\today}

\begin{abstract}
Rare-earth ions in crystals are a proven solid-state platform for quantum technologies in the ensemble regime and attractive for new opportunities at the single ion level. Among the trivalent rare earths,  ${}^{171}\mathrm{Yb}^{3+}$ is unique in that it possesses a single 4f excited-state manifold and is the only paramagnetic isotope with a nuclear spin of 1/2. In this work, we present measurements of the optical and spin properties of $^{171}$Yb$^{3+}$:YVO$_4$~to assess whether this distinct energy level structure can be harnessed for quantum interfaces. The material was found to possess large optical absorption compared to other rare-earth-doped crystals owing to the combination of narrow inhomogeneous broadening and a large transition oscillator strength. In moderate magnetic fields, we measure optical linewidths less than 3 kHz and nuclear spin linewidths less than 50 Hz. We characterize the excited-state hyperfine and Zeeman interactions in this system, which enables the engineering of a $\Lambda$-system and demonstration of all-optical coherent control over the nuclear spin ensemble. Given these properties, $^{171}$Yb$^{3+}$:YVO$_4$~has significant potential for building quantum interfaces such as ensemble-based memories, microwave-to-optical transducers, and optically addressable single rare-earth-ion spin qubits.
\end{abstract}

\maketitle{}

\section{Introduction}
Future quantum networks will incorporate a number of different quantum technologies, such as stationary qubits for high-fidelity logic operations and quantum memories for synchronization and long term storage \cite{Kimble2008,Northup2014,OBrien2009}. A successful network will require robust interfaces to coherently map quantum information between the best technologies, which may be based on disparate physical systems. For example, quantum transducers between the microwave and optical domains could be used to interface superconducting processors over long distances through optical networks \cite{Andrews2014}. 

Rare-earth ions (REIs) doped into crystalline hosts have demonstrated significant progress in implementing solid-state quantum technologies. REIs possess some of the longest optical and spin coherence lifetimes in the solid state \cite{Equall1994, Bottger2009a, MZhong2015, Rancic2017}, which has provided the foundation for numerous demonstrations of quantum memories and quantum interfaces \cite{Tittel2009, Hedges2010,  Sabooni2013, Saglamyurek2014, Gundogan2015, Laplane2016}. For interfaces involving both microwave and optical photons, REIs with an odd number of electrons (i.e. Kramers ions), such as erbium, neodymium, and ytterbium, are of interest due to their electron spin transitions. The large magnetic moments of these ions allow for strong interactions with microwaves, enabling fast operations and the potential for interfacing with superconducting qubits. Isotopes of these ions with non-zero nuclear spin also offer the possibility of long-term quantum storage \cite{Rancic2017}. This combination of properties creates the potential for building interfaces between microwave photons, optical photons, and long-lived nuclear spins. 

Among the Kramers ions, ytterbium is an attractive choice due to its simple level structure consisting of only two electronic multiplets. The optical transition between the lowest energy levels of these multiplets occurs around 980 nm, which is readily accessible by standard diode lasers. Furthermore, the\ybion isotope is unique among the trivalent REIs as the only Kramers ion with a nuclear spin of 1/2. This gives the simplest possible hyperfine energy structure allowing for both electron and nuclear spin degrees of freedom, reducing the complexity of optical preparation and manipulation of spin states \cite{Lauritzen2008, Baldit2010}. Recent work in \ybyso~\cite{Welinski2016a, Tiranov2017,  Lim2018, Ortu2017},  $\mathrm{Yb^{3+}\!:\!LiNbO_3}$ \cite{Kis2014}, and Yb$^{3+}$:YAG \cite{Bottger2016a} highlights the interest in this ion. In this work, we investigate\ybion doped into the host crystal \yvo. \yvo ~is an attractive choice for implementing quantum interfaces \cite{DeRiedmatten2008, TZhong2017} due to the ability to fabricate nanoscale devices \cite{TZhong2016} and high site symmetry in this material. Furthermore, previous work points to the potential for high oscillator strength transitions for Yb$^{3+}$ doped into \yvo \; \cite{Krankel2004}.

We present an initial survey of the properties of optical and nuclear spin transitions in \ybyvo at cryogenic temperatures. To determine whether this material can be used for efficient interactions with light, we characterized the strength and inhomogeneity of the optical transitions using high-resolution optical spectroscopy. Large hyperfine couplings and narrow optical inhomogeneous lines in this material result in resolved optical transitions between the hyperfine states, which allowed for characterization of the excited-state spin Hamiltonian directly from absorption measurements in an applied magnetic field. Knowledge of the spin Hamiltonian enables the identification of magnetic field orientations that create strongly spin-conserving transitions (for cyclic transitions) or strongly spin-mixing optical transitions (allowing for efficient lambda systems). To assess the possibility of storage and manipulation of quantum information in this material, we measured the coherence properties of the optical and nuclear spin transitions as a function of applied magnetic field. To demonstrate the potential for all-optical control of the nuclear spin states, we also measured spin echoes using bichromatic Raman pulses. 

This article is organized as follows: \sect{background} presents the material properties of the samples used in this work and the spin Hamiltonian used to model this system. \sect{experiment} describes the experimental methods and apparatus. \sect{results} presents the experimental results and discussion. This section is further divided into subsections: A (Optical absorption spectroscopy), B (Optical transition strengths), C (Excited-state lifetime), D (Optical coherence measurements), E (Nuclear spin measurements), and F (All-optical spin coherence measurements).

\section{Background}
\label{sec:background}
\subsection{Material properties}

\yvo~(also called yttrium orthovanadate or YVO) forms a zircon tetragonal crystal with $D_{4h}$ symmetry \cite{Wyckoff1963}. Ytterbium substitutes for yttrium in sites of local $D_{2d}$ point group symmetry. The $z$-axis of the site coincides with the crystalline 4-fold axis (the \textit{c}-axis of the crystal). The uniaxial nature of this site reduces the number of parameters needed to characterize the system compared to a lower symmetry crystal such as \yso \cite{Welinski2016a,Tiranov2017}. 

The majority of measurements presented in this paper were performed in samples cut from a boule of \yvo ~doped with isotopically enriched \ybion custom grown by Gamdan Optics. The concentration of ${}^{171}\mathrm{Yb}$ was measured to be 100 ppm using secondary ion mass spectrometry (SIMS). The samples were cut and polished to various thicknesses appropriate to each measurement. Fluorescence lifetime measurements were performed using a nominally undoped sample of \yvo~(Gamdan Optics), which was measured using SIMS to have a residual \ybion concentration of approximately 2 ppm.

\subsection{Spin Hamiltonian}
%\label{sec:spinhamiltonian}

%This section outlines the effective spin Hamiltonian used to describe the energy level structure of \ybion in \yvo. 

\begin{figure}[t!]
 % \centering
  \includegraphics[height = 0.25\textheight]{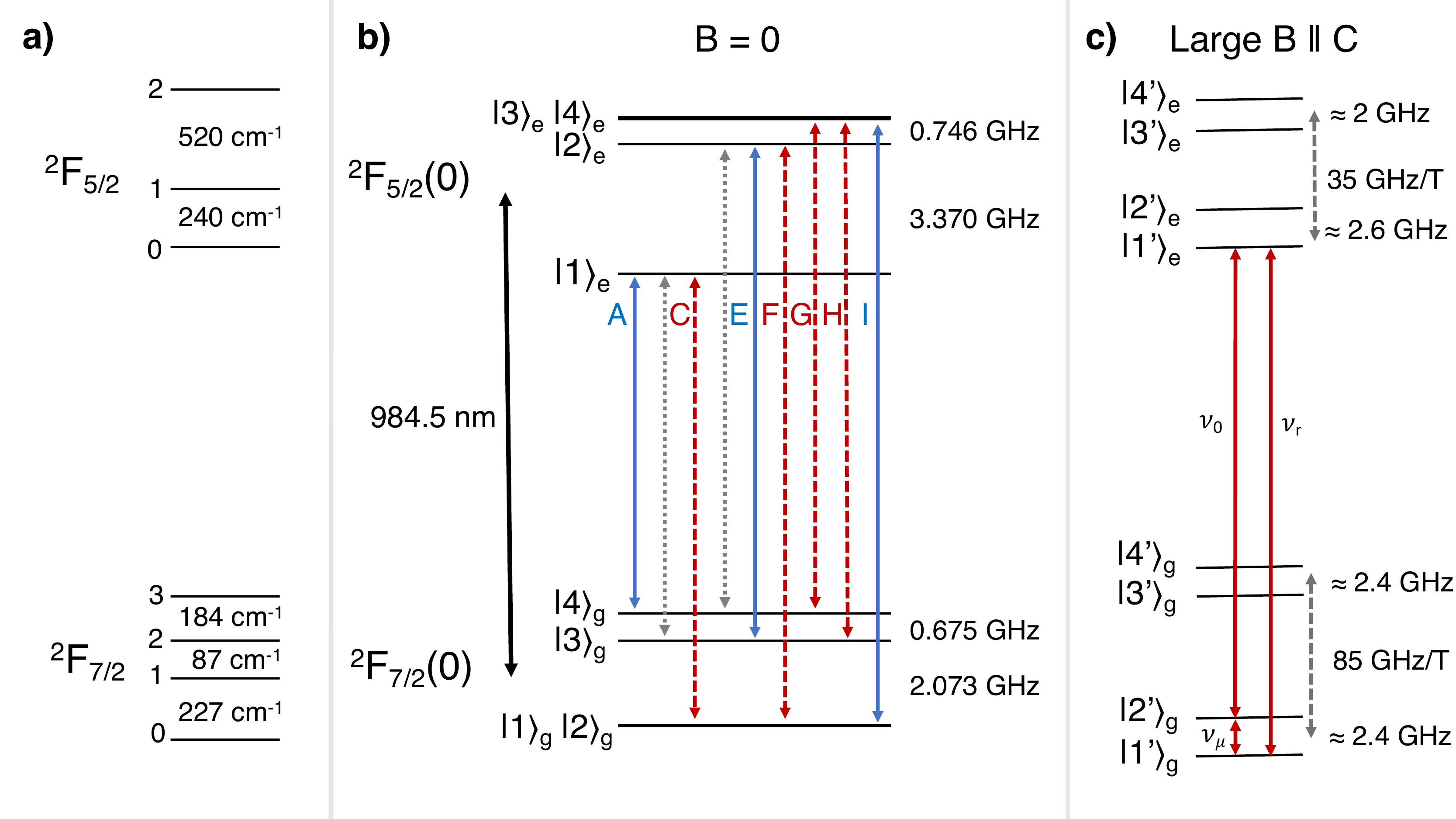}
\caption{Energy level diagram for \ybyvo. a) Crystal field splittings of \ybyvo reproduced from \cite{Pestryakov2005} b) Zero-field energy level diagram for the ${}^2F_{7/2}(0) \rightarrow {}^{2}F_{5/2}(0)$ transition of \ybyvo at 984.5 nm studied in this paper. Energy splittings in the ground and excited state are extracted from the excited-state hyperfine tensor determined in this work and previous measurements of the ground-state hyperfine tensor \cite{Ranon1968}. The transitions corresponding to the observed absorption spectrum in \fig{zerofieldabs} for $E \parallel c$ ($ E \perp$ c) are shown in solid blue (dashed red). The dotted grey lines correspond to transitions that are forbidden by symmetry. c) Energy level diagram for the linear Zeeman regime with $B\parallel c$ with arrows denoting the transitions studied in this work.}
\label{fig:energydiagram}
\end{figure}

The $4f^{13}$ configuration of $\mathrm{Yb}^{3+}$ consists of two electronic multiplets: ${}^2F_{7/2}$ and ${}^{2}F_{5/2}$. In the crystal field of \yvo, the ground-state multiplet (${}^2F_{7/2}$) splits into four Kramers doublets and the excited-state multiplet (${}^{2}F_{5/2}$) splits into three Kramers doublets. The energies of these crystal-field levels have been measured previously \cite{Pestryakov2005} and are shown in \fig{energydiagram}. At liquid helium temperatures, only the lowest energy doublet of the ground state is thermally occupied. The optical transition of interest for quantum interfaces is between the lowest energy doublets of the ground state and excited state (${}^2F_{7/2}(0) \rightarrow {}^{2}F_{5/2}(0)$). This transition occurs at approximately 984.5 nm for $\mathrm{Yb}^{3+}$ doped into \yvo . 

In this work, we focus on the ${}^{171}\mathrm{Yb}$ isotope, which has a nuclear spin $I = 1/2$. Treating the Kramers doublets as effective spins with $S = 1/2$, we can describe the system with the following spin Hamiltonian \cite{Abragam1970}:  
\beq
 H_{\textrm{eff}} = \mu_B\mathbf{B} \cdot \mathbf{g} \cdot \mathbf{S} + \mathbf{I} \cdot \mathbf{A} \cdot \mathbf{S} - \mu_{n}  \mathbf{B} \cdot \bf{g_n} \cdot \bf{I}.
 \label{eq:spinham}
 \eeq 
 
The first term is due to the electronic Zeeman interaction, where $\mu_B$ is the Bohr magneton, $\bf{B}$ is the applied magnetic field, $\bf{g}$ is the electronic Zeeman tensor, and $\bf{S}$ is the spin 1/2 operator. The second term describes the coupling between the electron spin and nuclear spin via the hyperfine interaction, where $\bf{I}$ is the nuclear spin operator and $\bf{A}$ is the hyperfine interaction tensor. The last term arises from the nuclear Zeeman interaction, where $\mu_n$ is the nuclear magneton and $\bf{g_n}$ is the nuclear Zeeman tensor. For $^{171}\mathrm{Yb}$ in \yvo, the non-zero components of $\mathbf{g_n}$ will be of the order of the gyromagnetic moment of the free nucleus $g_n=0.987$, which leads to a nuclear Zeeman interaction $\sim 2000$ times smaller than the electronic Zeeman term. For the magnetic field values used in this work, we treat this interaction by incorporating it into the electronic Zeeman tensor. 

The energy structure in the absence of an external magnetic field ($\mathbf{B} = 0$) is determined by the hyperfine interaction $\mathbf{I} \cdot \mathbf{A} \cdot \mathbf{S} $. In the site symmetry of \yvo, the degeneracy of these levels is partially lifted and the Hamiltonian has the following eigenvalues at zero field: $E = \frac{A_\parallel}{4} , \frac{A_\parallel}{4} , \frac{-A_\parallel + 2 A_\perp}{4} , \frac{-A_\parallel - 2 A_\perp}{4} $, 
%\begin{align}
%E_{1,2} &= \frac{A_\parallel}{4} \\
%E_{3,4} &= \frac{-A_\parallel \pm 2 A_\perp}{4} 
%\end{align}
where $A_\perp$ and $A_\parallel$ are the components of the hyperfine tensor $\bf{A}$ perpendicular and parallel to the crystal symmetry axis (the \textit{c}-axis) \cite{Abragam1970}. The order of the energies is determined by the signs of these components, which we have determined to be $A^g_{\parallel}<0$ and $A^g_{\perp},A^e_{\parallel},A^e_{\perp}>0$ (see Section \ref{subsec:opticalabsorption}) with the superscript $g\, (e)$ denoting the ground (excited) state. The corresponding eigenstates numbered from lowest to highest energy are 
\begin{align}
\ket{1}_g  &=  \ket{ \uparrow \Uparrow }_g & \ket{1}_e &= \frac{1}{\sqrt{2}} \left(\ket{ \uparrow \Downarrow }_e - \ket{ \downarrow \Uparrow }_e \right) \\
\ket{2}_g &=  \ket{ \downarrow \Downarrow }_g & \ket{2}_e &= \frac{1}{\sqrt{2}} \left(\ket{ \uparrow \Downarrow }_e + \ket{ \downarrow \Uparrow }_e \right)\\
\ket{3}_g &= \frac{1}{\sqrt{2}} \left(\ket{ \uparrow \Downarrow }_g - \ket{ \downarrow \Uparrow }_g \right) & \ket{3}_e &= \ket{ \uparrow \Uparrow }_e  \\
\ket{4}_g &= \frac{1}{\sqrt{2}} \left(\ket{ \uparrow \Downarrow }_g + \ket{ \downarrow \Uparrow }_g \right) & \ket{4}_e  &= \ket{ \downarrow \Downarrow }_e.
\end{align}
We denote the electron spin components as $\ket{\uparrow} \equiv \ket{S_z = \frac{1}{2}}$, $\ket{\downarrow} \equiv \ket{S_z = - \frac{1}{2}}$  and the nuclear spin components as $\ket{\Uparrow} \equiv \ket{I_z = \frac{1}{2}}$, $\ket{\Downarrow} \equiv \ket{I_z = -\frac{1}{2}}$. 

For high magnetic fields applied along the \textit{c}-axis, the electronic Zeeman interaction dominates over the hyperfine interaction. In this regime, mixing between the electron and nuclear spin is greatly reduced and the states effectively become
\begin{align}
\ket{1'}_g  & \approx \ket{ \uparrow \Uparrow } & \ket{1'}_e  &\approx \ket{ \downarrow \Uparrow }\\
\ket{2'}_g  & \approx \ket{ \uparrow \Downarrow } & \ket{2'}_e &\approx \ket{ \downarrow \Downarrow } \\
\ket{3'}_g  & \approx \ket{ \downarrow \Downarrow } & \ket{3'}_e &\approx \ket{ \uparrow \Downarrow }\\
\ket{4'}_g  & \approx \ket{ \downarrow \Uparrow }  & \ket{4'}_e &\approx \ket{ \uparrow \Uparrow }.
\end{align}
We have again numbered the states from lowest to highest energy using the fact that $g_\parallel<0$  for the ground state and  $g_\parallel>0$ for the excited state (see Section \ref{subsec:opticalabsorption}). The prime is used to distinguish between the high field and zero field state labels. In this work, we focus on the coherence properties of the optical and nuclear spin transitions in the regime where the linear Zeeman interaction is dominant.

\section{Experimental Methods}
\label{sec:experiment}
High-resolution laser absorption scans with a home-built external-cavity diode laser (ECDL) using the design from \cite{Cook2012}~were performed to measure the inhomogeneous linewidth and absorption of the $^2F_{7/2}(0) \rightarrow {}^{2}F_{5/2}(0)$ transition. The energies of the optical transitions were extracted from absorption scans taken with magnetic fields applied along the crystal symmetry axes and used to determine the excited-state spin Hamiltonian. For this purpose, we used a $90 \; \mu \mathrm{m}$ thick a-cut sample of \ybyvo. This thickness was chosen such that the sample was not overabsorbing at 2 K. The sample was mounted in a custom sample mount and masked to avoid spurious light leakage around the crystal that could lead to inaccurate measurements of the optical depth. For the data presented here, the probe light propagated parallel to the \textit{a}-axis of the crystal and perpendicular to the applied magnetic field ($\mathbf{k} \perp \mathbf{B}, c$). Additional axial spectra ($\mathbf{k}\parallel c$) were taken to confirm the electric dipole nature of the optical transitions \cite{Liu2005}. The absorption was determined by measuring the transmission of the ECDL on a photodetector (New Focus 2031) as the frequency of the laser was scanned across resonance. The center frequency of the scan was calibrated with a wavemeter (Burleigh WA-1500) and the frequency detuning of the scan was calibrated using a Fabry-Perot reference cavity. The absorption experiments were performed in an Oxford Spectromag cryostat at a temperature of 2 K with an applied magnetic field of up to 6 T. 

For measurements of the excited-state, optical coherence, and spin coherence lifetimes, the optical transitions were addressed using a single frequency Ti:Sapphire laser (M-Squared Solstis) that was gated by an 80 MHz acousto-optic modulator in a double-pass configuration to create the required pulse sequence. For measurements of the coherence properties and inhomogeneity of the nuclear spin transition, the nuclear spin transition was addressed directly using a coaxial transmission line mounted directly next to the sample. 

The excited-state lifetime was measured from the time-resolved fluorescence decay. We performed pulsed excitation on the  $^2F_{7/2}(0) \rightarrow  {}^2F_{5/2}(0)$ transition and collected the resulting fluorescence to the upper crystal field levels of the ground state (i.e. $^2F_{5/2}(0) \rightarrow  {}^2F_{7/2}(1-3)$) using a 1000 nm long-pass filter. The fluorescence counts as a function of time were recorded using a silicon APD (Perkin-Elmer). Fluorescence measurements were performed in a $500\; \mathrm{\mu m}$ thick sample that was nominally undoped (residual\ybion concentration of $\sim 2$ ppm) and a $200\; \mathrm{\mu m}$ thick 100 ppm sample of \ybyvo\!. These measurements were performed at 4 K with zero applied magnetic field in a Montana Instruments cryostat using a home-built confocal microscope setup.

The coherence properties of the optical transition were investigated using two-pulse photon echo decays as a function of magnetic field strength. For this purpose, two-pulse photon echoes on the \opttransvr~transition were measured using heterodyne detection.  During the echo sequence, a fiber-based phase modulator EOM (IXBlue NIR-MPX-LN-20) was driven by a microwave source (Windfreak Synth HD) at 500 MHz to create an optical sideband resonant with the optical transition. The resulting echo was detected as a beat at the sideband frequency using an InGaAs photodiode (Thorlabs DET08CFC, BW = 5 GHz). Typical $\pi$-pulses for this measurement were $4 \; \mu s$ long. 

Optical coherence measurements were performed with the sample mounted on the still plate of a Bluefors dilution refrigerator at a temperature of 650 mK. These measurements used a 500 $\mu\mathrm{m}$ thick $c$-cut 100 ppm sample with $\mathbf{k} \parallel c$.  The light entered the refrigerator via single-mode optical fiber and was focused onto the back surface of the sample, which was coated in gold to enhance reflection. The reflected light was coupled back into the fiber and directed to the photodetector with a fiber beam splitter. A variable magnetic field of up to 1.5 T was applied along the crystal \textit{c}-axis using a homebuilt superconducting solenoid.

The inhomogeneous linewidth of the nuclear spin transition was measured using continuous-wave Raman heterodyne detection \cite{Wei1996}. Frequency-swept microwave tones from a tracking generator (Anritsu) were amplified and applied to the sample. The coherence generated on the \nucspintrans~nuclear spin transition was mapped to an optical coherence by applying a continuous wave laser to the \opttransvo~optical transition at frequency $\nu_0$, which resulted in coherent Raman scattering on the \opttransvr~optical transition at $\nu_r$. This signal was detected on the transmitted optical beam as a beat at the microwave transition frequency ($\nu_0 - \nu_r$) using an InGaAs photodiode.

The nuclear spin coherence was measured using two-pulse spin echoes. Coherent manipulation on the nuclear spin state was performed with both direct microwave excitation and all-optical excitation with bichromatic Raman pulses \cite{Blasberg1994, Blasberg1995, Walther2016, Serrano2017}. The ions were first initialized into the $\ket{1'}_g$ state via optical pumping on the \opttransvo~transition. For direct manipulation, the echo sequence was performed using tones generated by a microwave source tuned to the \nucspintrans~nuclear spin transition. Pulses were generated using microwave switches (Minicircuits ZASWA-2-50DR+) with typical $\pi$ pulse lengths of $100 \; \mu s$. For all-optical spin echoes, the nuclear spin transition was coherently manipulated via the shared excited state $\ket{1'}_e$ by applying bichromatic pulses to the \opttransvo~and \opttransvr~transitions as depicted in \fig{energydiagram}c. Typical spin $\pi$ pulses for the all-optical sequence were $8 \; \mu s$. The two optical frequencies were generated by driving a fiber-based phase modulator with a microwave source tuned to the nuclear spin transition frequency. The relative power of the two optical frequencies was chosen to maximize the echo signal. The resulting spin echo was optically detected via Raman heterodyne scattering by applying a readout pulse to the \opttransvo~transition at the time of the echo. The signal was detected as a beat on the probe laser at the nuclear spin transition frequency.% using an InGaAs photodiode (Thorlabs DET08CFC, BW = 5 GHz ).
%, amplified, mixed down to 10 MHz, and recorded with a digital oscilloscope. 

Nuclear spin coherence measurements were performed at approximately 700 mK. These measurement were done in transmission through a 2 mm thick a-cut sample with $\mathbf{k} \perp c, \mathbf{B}$. The polarization of the input light was set using a fiber polarization controller to maximize the echo signal. A variable magnetic field was applied to the crystal using a set of homebuilt superconducting Helmholtz coils. For the direct microwave measurements, the magnetic field was applied along the \textit{c}-axis. For the all-optical measurements, the magnetic field was applied $\sim20^\circ$ from the \textit{c}-axis. As described in \sect{results}, this was done to help equalize the strengths of the optical transitions used in the measurement.

%%%%%%%%%%%%%%%%%%
%%%%%%%%%%%%%%%%%%
\section{Optical and spin properties of ${}^{171}\mathbf{Y\MakeLowercase{b}:YVO_4}$}
\label{sec:results}
\subsection{Optical absorption spectroscopy}
\label{subsec:opticalabsorption}

%In this section, we present high-resolution absorption measurements for the $^2F_{7/2}(0) \rightarrow  ^2F_{5/2}(0)$ transition. From these measurements, we extract absorption coefficients, inhomogeneous linewidths, and the excited state Zeeman and hyperfine tensors.

\begin{figure}[t!]
  \centering
  \includegraphics[height = 0.25\textheight]{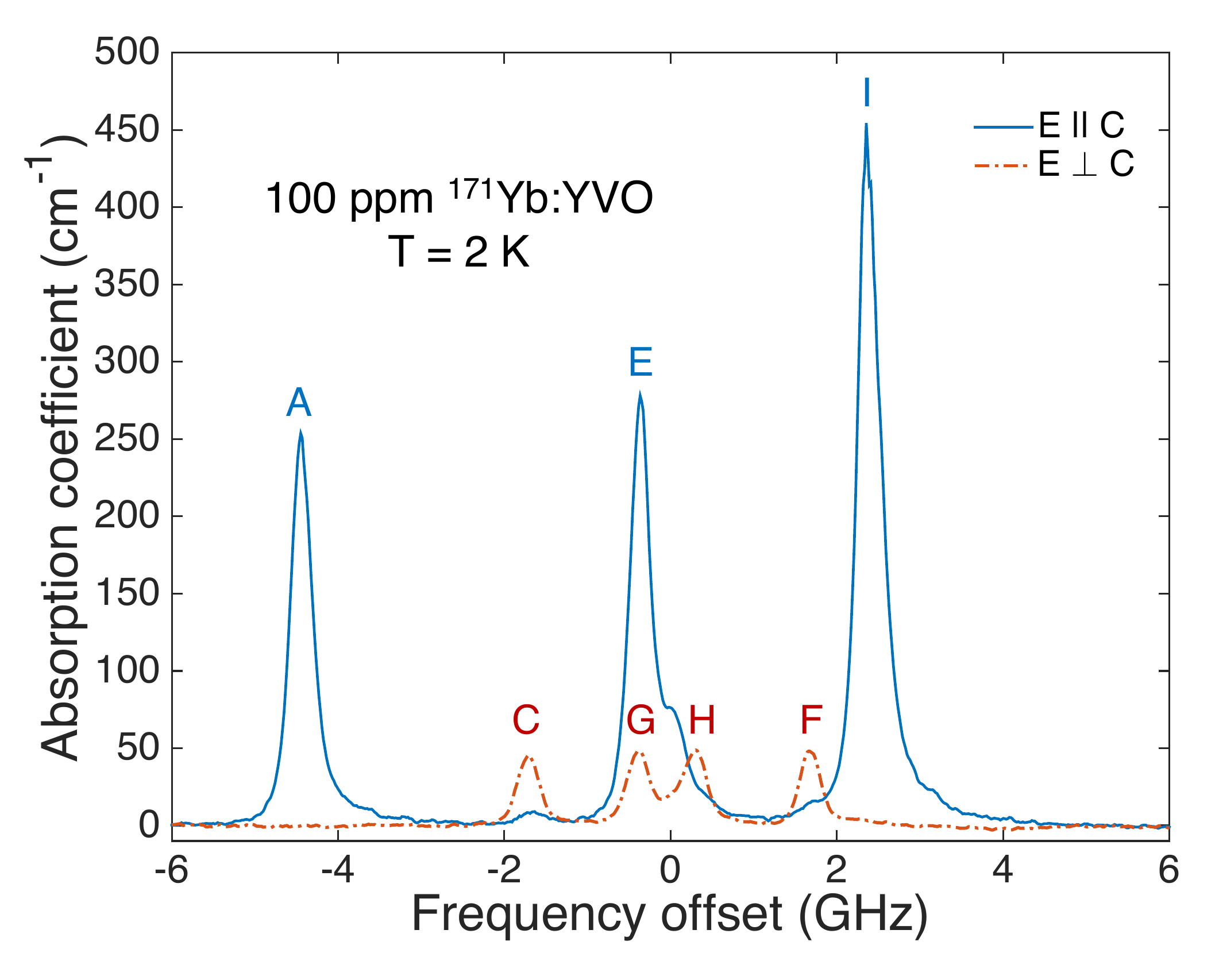}
\caption{Optical absorption spectra of the $^2F_{7/2}(0) \rightarrow  {}^2F_{5/2}(0)$ transition of \ybyvo at 2 K and zero applied magnetic field for light polarized parallel (solid blue) and perpendicular (dashed red) to the crystal \textit{c}-axis. }
\label{fig:zerofieldabs}
\end{figure}

The zero-field absorption spectra for the $^2F_{7/2}(0) \rightarrow  {}^2F_{5/2}(0)$ transition of \ybyvo at 2 K is shown in \fig{zerofieldabs}. We observed narrow inhomogeneous linewidths (average FWHM = 275 MHz), which allowed us to resolve and address individual optical-hyperfine transitions. For $\mathbf{E} \parallel c$, we observed three resolved transitions with a peak absorption coefficient for the strongest transition of 450 $\mathrm{cm}^{-1}$. For $\mathbf{E}\perp c$, we observed four resolved transitions with a peak absorption of 50 $\mathrm{cm}^{-1}$. The corresponding transitions on the energy diagram are labeled in \fig{energydiagram}. The strong polarization selection rules between the optical hyperfine transitions observed in \fig{zerofieldabs} are consistent with those derived for electric-dipole transitions based on the site's point group symmetry \cite{Koster1963}. 

 %In terms of the effective spin Hamiltonian, the selection rules at zero field can be written as $\Delta m_e = 0$ for $\mathbf{E} \parallel c$ and $\Delta m_e = \pm 1$ for $\mathbf{E} \perp c$  with $\Delta m_I = 0 $ \cite{?}. These selection rules correctly predict the observed behavior. 

We observed a peak at zero detuning, which corresponds to the presence of zero spin isotope in the sample (measured to be $<10$ ppm from SIMS). We also noted the presence of additional satellite lines due to the 173 isotope. 

The ground state Zeeman and hyperfine tensors of \ybyvo have been determined using EPR \cite{Ranon1968}, so a description of the system requires finding the corresponding values for the excited state. For a uniaxial crystal, this reduces to four parameters: the components of $\bf{g}$ and $\bf{A}$, parallel and perpendicular to the crystal symmetry axis. The values for $\bf{A}$ can be determined by the energy level structure in the absence of a magnetic field, while $\bf{g}$ can be determined from the energy level structure as magnetic fields are applied parallel and perpendicular to the crystal's $c$-axis.

Fitting to the energy level splittings extracted from the absorption spectra, we find agreement with previously published data for the ground-state $\bf{A}$ tensor \cite{Ranon1968} ($A^g_\parallel/h = -4.82\;\mathrm{GHz}$, $A^g_\perp/h = 0.675\;\mathrm{GHz}$) and we determine the principal values of the excited-state hyperfine tensor to be $A^e_\parallel/h = 4.86 \pm 0.05 \;\mathrm{GHz}$ and $A^e_\perp / h = 3.37 \pm 0.05 \;\mathrm{GHz}$. 

\begin{figure}[t!]
  \centering
\includegraphics[height = 0.25\textheight]{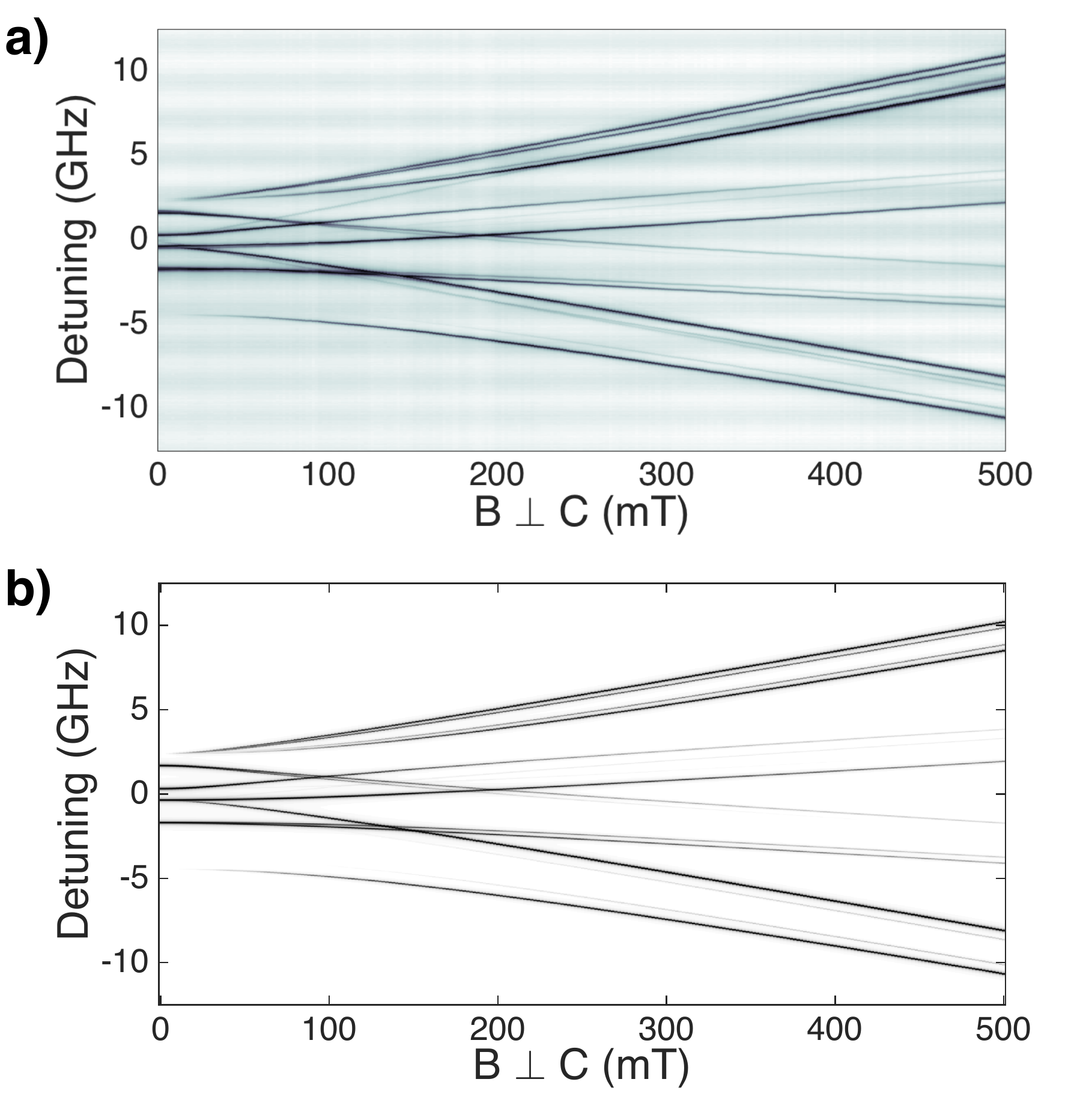}
\caption{a) Typical high-resolution absorption spectra for a magnetic field ramp with $B , \mathbf{k} \perp c$ and $ \mathbf{E} \perp c$ showing resolved optical hyperfine transitions. Darker regions correspond to higher absorption. b) Simulated absorption spectra using the experimentally determined spin Hamiltonian for magnetic field ramp with $B , \mathbf{k} \perp c$ and $ \mathbf{E}  \perp c$. }
\label{fig:magnetramps}
\end{figure}

The excited-state $\bf{g}$ tensor was determined by measuring the frequency of the optical transitions with magnetic fields applied parallel and perpendicular to the crystal's $c$-axis. \fig{magnetramps}a shows an example of one such measurement in which the absorption was recorded while the magnetic field perpendicular to the crystal \textit{c}-axis was continuously ramped. By fitting to the energy levels extracted from this spectra and similar measurements for other field orientations, we determine $g_{e,\parallel} = 2.51 \pm 0.1$ and $g_{e,\perp} = 1.7 \pm 0.1$.  \fig{magnetramps}b shows the absorption spectra expected from the spin Hamiltonian, which we see enables accurate predictions of the energy level splittings and relative transition absorption oscillator strengths in this case. 

\subsection{Optical transition strengths}
\label{subsec:transitionstrengths}
The strength of the optical transitions can be characterized by assigning an oscillator strength to each individual transition. The absorption oscillator strength for a transition $\ket{i} \rightarrow \ket{j}$ for a polarized spectrum is given by \cite{DiBartolo1968,Henderson2006}
\beq
f_{ij} = 4 \pi \epsilon_0 \frac{m_e c}{\pi e^2}\frac{1}{N} \sum_i\frac{9 n_i}{(n_i^2 + 2)^2}\int \alpha_i(\nu) d\nu, 
\eeq
where $\epsilon$ is the vacuum permittivity, $m_e$ is the mass of the electron, $e$  is the charge on the electron, $c$ is the speed of light, N is the number density, and the summation is over tmhe three orthogonal polarizations states with $\alpha_i$ and $n_i$ absorption coefficient and index of refraction, respectively. For \yvo, $n_\parallel = 2.17$ and $n_\perp = 1.96$ at 984 nm \cite{Shi2001}.  

Assuming a doping density of 100 ppm, the number density of $\mathrm{Yb}^{3+}$ in \yvo~is calculated to be $N = 1.24 \times 10^{18} \; \mathrm{cm^{-3}}$, which is distributed between the four ground state levels according to Boltzmann statistics at 2 K. The integrated absorption coefficient and corresponding oscillator strengths for the observed transitions are summarized in Table \ref{oscillatorstrengths}. We measure an average oscillator strength of $5.3 \times 10^{-6}$ for transitions allowed for $\mathbf{E}\parallel c$ (transitions A, E, I in \fig{zerofieldabs}) and $1.8 \times 10^{-6}$ for transitions allowed for $\mathbf{E}\perp c$ (transitions C, F, G, H in \fig{zerofieldabs}). 

%(Need to make a note on how the degeneracy factors into these calculations. ``Useful oscillator strength") 

\begin{table}[htp]
\caption{Absorption properties of the \ybyvo transitions as labeled in \fig{zerofieldabs}, including the transition polarization \cite{Koster1963}, integrated absorption coefficient, oscillator strength, and radiative decay rate at zero magnetic field.}
\begin{center}
\begin{tabular}{| c | c | c | c | c |}
\hline
Trans. & Pol.& $\int\alpha(\nu) d \nu$ $ \mathrm{(GHz/cm)}$ & $f \;(10^{-6})$ & $1/\tau_{\mathrm{rad}} \; (\mathrm{kHz}) $\\
\hline
%A & $\pi$ & 97.3 & 1.78 & 0.75\\
%C & $\sigma$ &16.4 &0.334 &  2.92\\
%E & $\pi$ & 102.7 & 1.85 & 0.73\\
%F & $\sigma$ & 17.4 & 0.353 & 2.75\\
%G & $\sigma$ & 20.2 & 0.879 & 4.43\\
%H & $\sigma$ & 19.9& 0.851 & 4.58\\
%I & $\pi$ & 189.7 & 1.622 & 0.83\\
A & $\pi$ & 97.3 & 5.4 & 1.3\\
C & $\sigma$ &16.4 & 1.0 &  0.3\\
E & $\pi$ & 102.7 & 5.5 & 1.4\\
F & $\sigma$ & 17.4 & 1.1 & 0.4\\
G & $\sigma$ & 20.2 & 2.6 & 0.2\\
H & $\sigma$ & 19.9& 2.6 & 0.2\\
I & $\pi$ & 189.7 & 4.9 & 1.2\\
\hline

\end{tabular}
\end{center}
\label{oscillatorstrengths}
\end{table}%

The radiative lifetime for the $^2F_{5/2}(0) \rightarrow  {}^2F_{7/2}(0)$ transitions can be determined from the absorption measurements. The radiative lifetime for a transition $\ket{j} \rightarrow \ket{i}$ is related to the oscillator strength by \cite{DiBartolo1968,Henderson2006}
%\beq
%\frac{1}{\tau_{rad}} = \frac{1}{1.5 \times 10^4} \frac{(n^2 + 2)^2 }{9 n} \frac{n^2}{\lambda_0^2} f_{ji},
%\eeq
\beq
\frac{1}{\tau_{rad}} = \frac{2 \pi e^2}{\epsilon_0 m_e c} \frac{(n^2 + 2)^2 }{9 n} \frac{n^2}{\lambda_0^2} \frac{f_{ji}}{3}, %\approx  \frac{1}{1.5\times 10^4} \mathrm{\frac{m^2}{s}} \frac{(n^2 + 2)^2 }{9 n} \frac{n^2}{\lambda_0^2} \frac{f_{ji}}{3},
\eeq
where $n$ is the index of refraction, $\lambda_0$ is the wavelength in vacuum, and $f_{ji}$ is the emission oscillator strength. The emission oscillator strength is related to the absorption oscillator strength $f_{ij}$ by $f_{ji} = \frac{g_i}{g_j}f_{ij}$, where $g_i$ ($g_j$) is the degeneracy of state $\ket{i}$ ($\ket{j}$). The calculated emission rates for the observed transitions are included in Table \ref{oscillatorstrengths}. From these rates, we obtain an average radiative rate of $1/\tau_{rad} = 1/(590 \; \mathrm{\mu s})$ for the $^2F_{5/2}(0) \rightarrow  {}^2F_{7/2}(0)$ transitions. 

%%%%%%%%%%%%%%%%%
%%%%%%%%%%%%%%%%%
\subsection{Excited state lifetimes}
\label{subsec:excitedstate}
%\textbf{Sentence on the excited state lifetime. Also sentence explaining difference between radiative lifetime and excited state lifetime.}
The excited-state lifetime is important for optical preparation of population among the spin states and sets the upper limit on the optical coherence time. The measured excited-state lifetime allow us to determine the optical branching ratio between the crystal field levels, which is important in the context of Purcell enhancement in nanophotonic cavities \cite{Purcell1946}. Here, we measure the excited-state lifetime through fluorescence decay.

\begin{figure}[t!]
  \centering
  \includegraphics[height = 0.25\textheight]{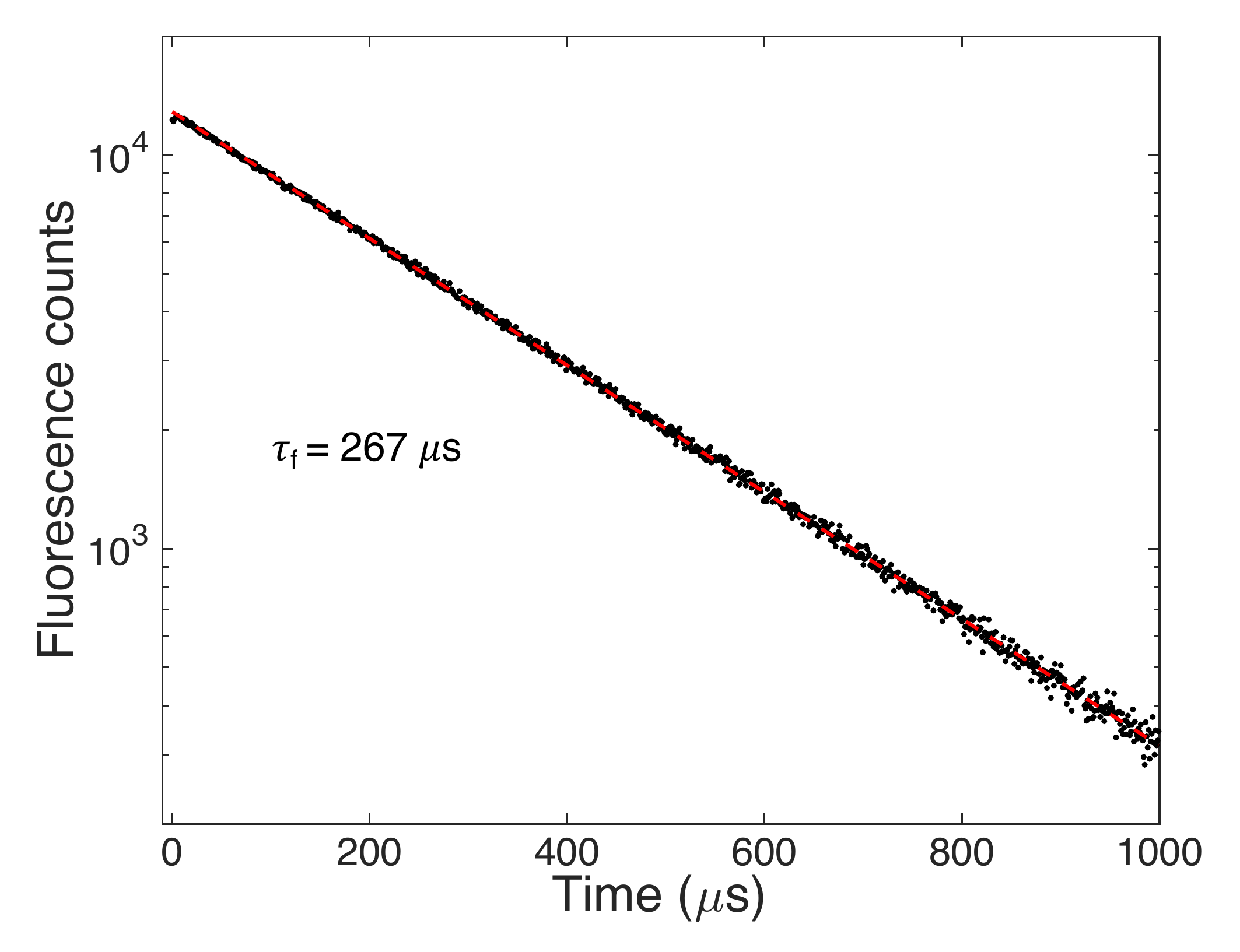}
\caption{Excited-state lifetime measurement via fluorescence decay. An exponential fit (dashed line) gives $\tau_f = 267 \pm 1 \; \mu s$.}
\label{fig:lifetime}
\end{figure}

To avoid the problem of radiation trapping \cite{Fan1994} observed in previous measurements of excited-state lifetimes in Yb-doped materials \cite{Bottger2016a,Krankel2004,Kisel2004}, the excited-state lifetime was measured in a nominally undoped sample of \yvo, which had a residual \ybion concentration of approximately 2 ppm. In this sample, we did not see variations in the optical lifetime within the inhomogeneous line or other signs of radiation trapping. A typical fluorescence decay in this sample is shown in \fig{lifetime}. Fitting to a single exponential gives a fluorescence lifetime of $ \tau_f = 267 \pm 1 \; \mathrm{\mu s}$. The branching ratio back to the same crystal field level ($^2F_{5/2}(0) \rightarrow  ^2F_{7/2}(0))$ is then given by $ \beta = \tau_f/\tau_{rad}$, where $\tau_f$  and $\tau_{rad}$ are the fluorescence and radiative lifetimes. Using the radiative lifetime obtained from the absorption measurements, we determine the branching ratio to be $\beta = 0.45$

We also note that in a $200 \; \mathrm{\mu m}$ thick 100 ppm sample we observed lifetimes longer than $500 \; \mathrm{\mu s}$ in the center of the inhomogeneous distribution that decreased to less than $300 \; \mathrm{\mu s}$ when the excitation pulse was detuned by 200 MHz from the center of the line. This behavior is attributed to radiation trapping due to the high optical depth and strong transition strengths of these ions.

%%%%%%%%%%%%%%%
\subsection{Optical coherence measurements}
\label{subsec:opticalcoherence}
To assess the ability to store quantum states in the material, we first investigate the coherence of the optical transition using two-pulse photon echoes (2PPE). For Kramers ions, we expect a dominant source of decoherence to be magnetic fluctuations due to magnetic dipole-dipole interactions between Yb ions \cite{Bottger2006}. One way to minimize this source of decoherence is to freeze out the electron spins by achieving a ground-state splitting much larger than $k_b T$ \cite{Bottger2006}. For \ybyvo\!\!, the energy-level splitting is maximized for a magnetic field along the crystal \textit{c}-axis. Here, we present measurements of the optical coherence in the linear Zeeman regime with the magnetic field along the \textit{c}-axis. While a comprehensive study is warranted to fully understand the decoherence mechanisms in this system, a large magnetic field applied parallel to \textit{c} provides insight on the maximum achievable coherence times in this material and the dominant decoherence mechanisms.

\begin{figure*}[t!]
  \centering
  \includegraphics[height = 0.25\textheight]{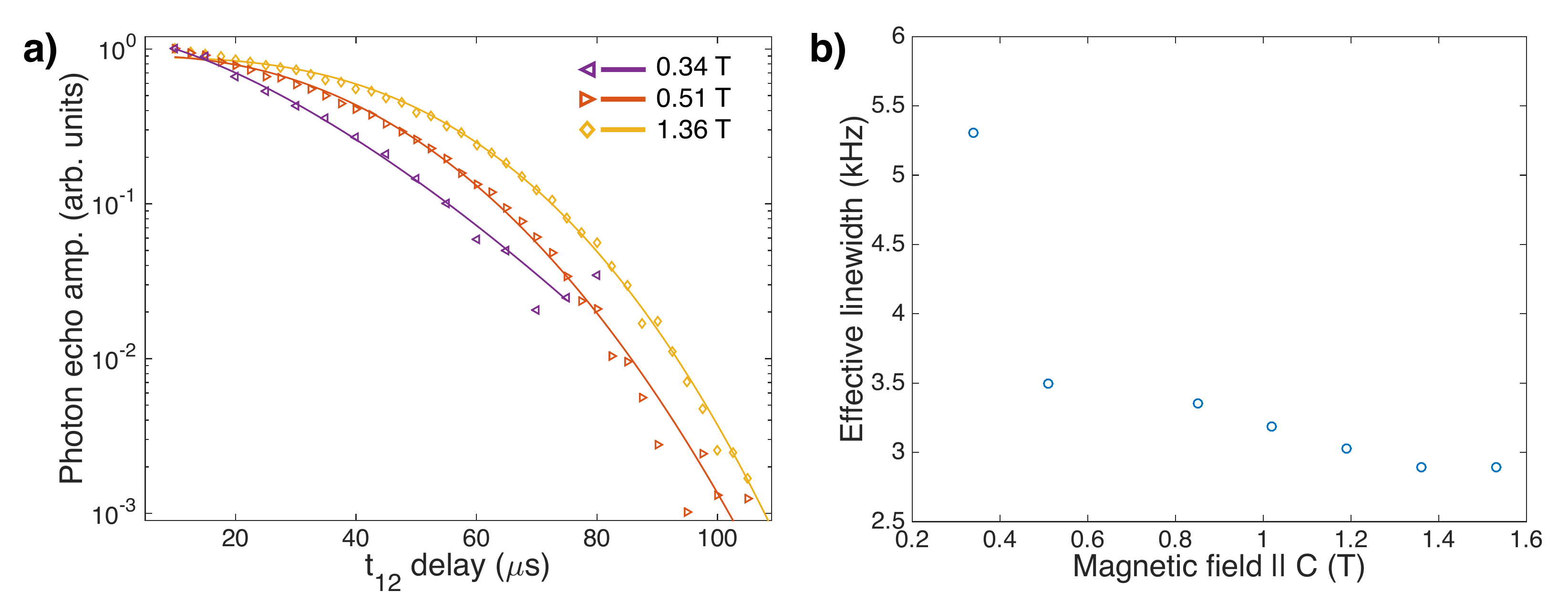}
\caption{a) Typical photon echo decays for $B \parallel c$ showing an increase of coherence time and strong non-exponential decays with increasing magnetic field. Fits to a Mims decay are shown for each field value as a solid line (see main text for details). b) Effective linewidth extracted from the fits to the photon echo decays for $B \parallel c$.}
\label{fig:opticalcoherences}
\end{figure*}

\fig{opticalcoherences}a shows typical photon echo decays for magnetic fields ranging from 340 mT to 1.36 T along the crystal \textit{c}-axis. We observed strong non-exponential decays, which can be attributed to spectral diffusion and described by a Mims decay \cite{Mims1968}. For heterodyne detection, the decay of the echo field is given by \cite{Mims1968}
\beq
E(t_{12}) = E_0 e^{-(2 t_{12}/T_m)^x},
\label{eq:mims}
\eeq
where $t_{12}$ is the delay between the two pulses used in the photon echo experiment, $x$ is the Mims parameter describing the spectral diffusion, and $T_m$ is phase memory time (the time at which the echo field amplitude decays to $e^{-1}$ of its initial value).  Fits to the Mims decay are shown as solid lines in \fig{opticalcoherences}a. From $T_m$, we can extract an effective homogeneous linewidth as $\Gamma_{\textrm{h,eff}} =(\pi T_m)^{-1}$. The effective linewidth as a function of applied magnetic field along the \textit{c}-axis is shown in \fig{opticalcoherences}b. 

We observed a decrease in the linewidth with applied magnetic field from $\sim\!5.5 \; \mathrm{kHz}$ at 340 mT to $\sim3\! \;\mathrm{kHz}$ at 1.5 T, which was the maximum magnetic field achievable for this measurement.  The reduction in linewidth for increasing magnetic field is expected for dephasing dominated by Yb-Yb spin flips and similar to that observed in other Kramers ions \cite{Bottger2006,Sun2002}. At the highest magnetic fields, we saw that the coherence no longer increased with applied field. The non-exponential decay and saturation of coherence time in the high-field limit are typical signs of the superhyperfine limit \cite{Ganem1991}. In this limit, magnetic fluctuations due to the electron spins are effectively frozen out and the main contribution to dephasing is interactions with the nuclei of the host material. 

We note that recent work in \ybyso \cite{Ortu2017} and \eryso \cite{Rakonjac2018a} demonstrated an increase in coherence time due to reduced sensitivity to magnetic fluctuations at zero first-order Zeeman (ZEFOZ) points at $\textbf{B} = 0$. While we did not explore the low-field regime in this work, \eq{spinham} predicts similar zero-field ZEFOZ transitions  in \ybyvo between levels $\ket{3}_g$ and $\ket{4}_g$ of the ground state and $\ket{1}_e$ and $\ket{2}_e$ of the excited state. 

%We also expect that applying a small magnetic field perpendicular to the \textit{c}-axis will lead to a similar coherence enhancement for transitions involving the other ground and excited states \cite{Ortu2017}. 

%%%%%%%%%%%%%%%
\subsection{Nuclear spin measurements}
\label{subsec:nuclearspincoherence}

The coherence times of the $^2F_{7/2}(0)$ nuclear spin transitions will determine the feasibility of long-term quantum information storage in this system. In this section, we present measurements on the inhomogeneous linewidth and coherence properties of the \nucspintrans~nuclear spin transition in the linear Zeeman regime.

\begin{figure}[t!]
  \centering
  \includegraphics[height = 0.25\textheight]{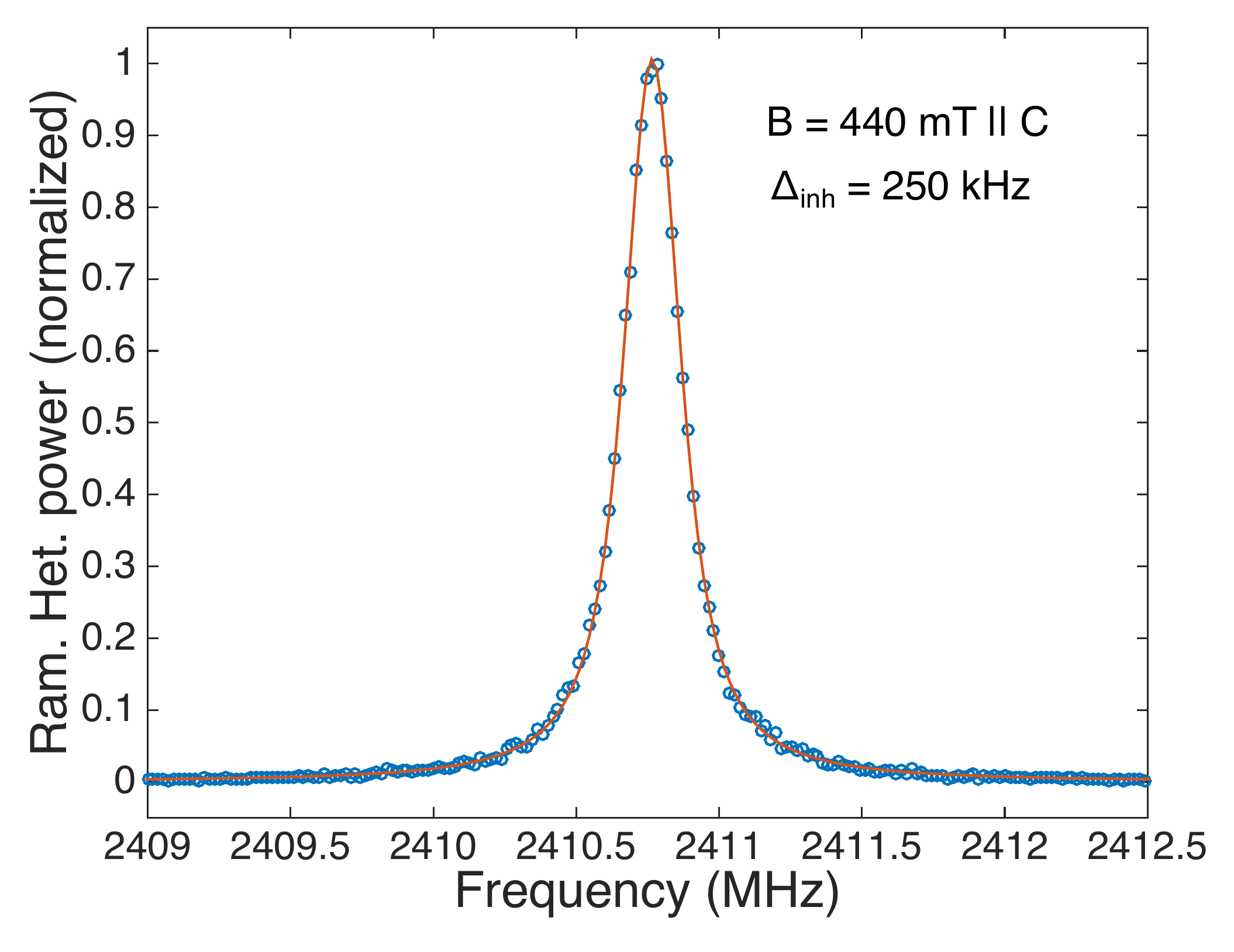}
\caption{Continuous-wave Raman heterodyne measurement of the nuclear spin inhomogeneity with B = 440 mT parallel to the \textit{c}-axis. A Lorentzian fit gives a FWHM of 250 kHz.}
\label{fig:spinCWramhet}
\end{figure}

The inhomogeneous broadening of the nuclear spin transition was measured using continuous-wave Raman heterodyne spectroscopy \cite{Wei1996}. \fig{spinCWramhet} shows a typical trace of the normalized Raman heterodyne signal power as the microwave frequency is swept across the resonance. Fitting this peak to a Lorentzian gives a FWHM of 250 kHz, which serves as an upper bound on the inhomogeneous broadening of the spin transition since the width of the observed signal can be power broadened by Rabi frequencies of the optical and microwave fields used in the measurement \cite{Fisk1990}. This measurement was done with a field of 440 mT along \textit{c}, but is typical of what was obtained for other magnetic field amplitudes applied along this direction. 

\begin{figure}[]
  \centering
  \includegraphics[height = 0.25\textheight]{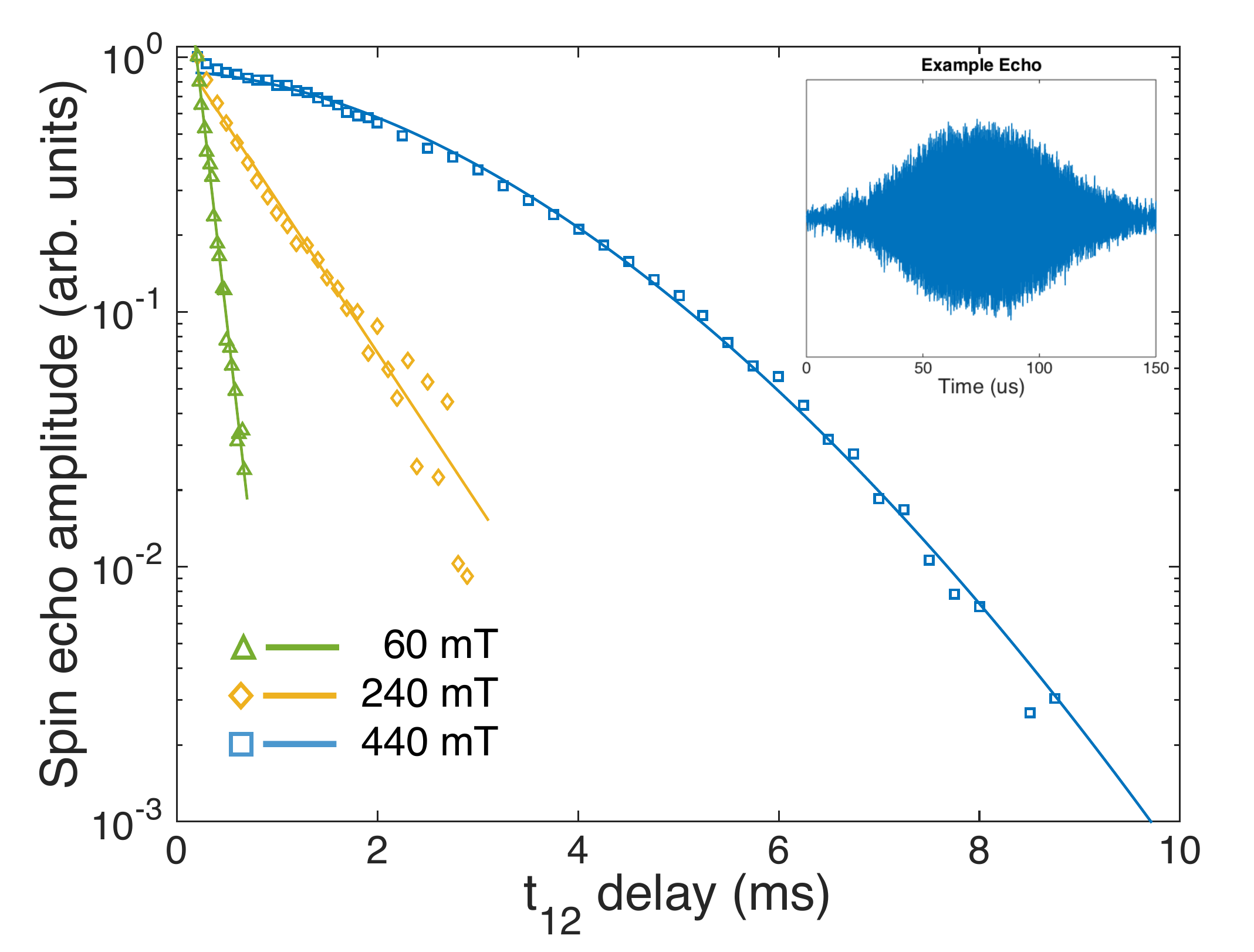}
\caption{Typical nuclear spin echo decays for increasing applied magnetic field along the \textit{c}-axis. The echo sequence is performed with direct microwave excitation and read out optically.}
\label{fig:spinT2_uw}
\end{figure}

The nuclear spin coherence was measured by spin echo decays with direct microwave manipulation of the \nucspintrans~spin transition and optical detection via coherent Raman scattering. \fig{spinT2_uw} shows typical nuclear spin echo decays for increasing magnetic fields along the \textit{c}-axis. For the higher field decays, we observed non-exponential decays resulting from time-varying dephasing mechanisms and again described by the Mims decay using \eq{mims}. We measured coherence times of 250 $\mathrm{\mu s}$ at 60 mT that increased up to 6.6 ms at a field of 440 mT, which was the maximum achievable magnetic field for the experimental configuration at the time of the measurement. Spectral hole burning measurements in this field configuration gave spin relaxation times  longer than $200$ ms, indicating that these coherence times are not lifetime limited. 

\subsection{All-optical spin coherence measurements}
\label{subsec:allopticalspin}

In addition to direct microwave excitation of the nuclear spins, we are interested in performing coherent all-optical control on the nuclear spins. All-optical control allows us to take advantage of relatively strong optical transitions to perform faster manipulations on the spin. This approach also removes the need for a microwave circuit to be incorporated next to the sample, which reduces the complexity of the experimental setup and prevents additional heating of the sample through the microwave excitation. As an initial demonstration of the potential for all-optical control in this system, we use an all-optical Raman echo technique \cite{Karlsson2017} to measure the coherence of the nuclear spin transition. The \nucspintrans~transition is addressed by applying bichromatic pulses to the lambda system formed by the \opttransvo~and \opttransvr~optical transitions. Efficient rephasing of coherence on the spin transition using bichromatic pulses in this fashion requires that the Rabi frequencies of the two transitions of the lambda system are equal \cite{Blasberg1995}. For magnetic fields applied parallel to the \textit{c}-axis, one transition of the lambda system is weak because the wavefunctions have approached separability and the optical transition cannot flip the nuclear spin. Moving the field off axis induces mixing of the nuclear and electronic states, which allows for a more favorable branching ratio between the two arms of the lambda system.

\begin{figure}[t!]
  \centering
  \includegraphics[height = 0.25\textheight]{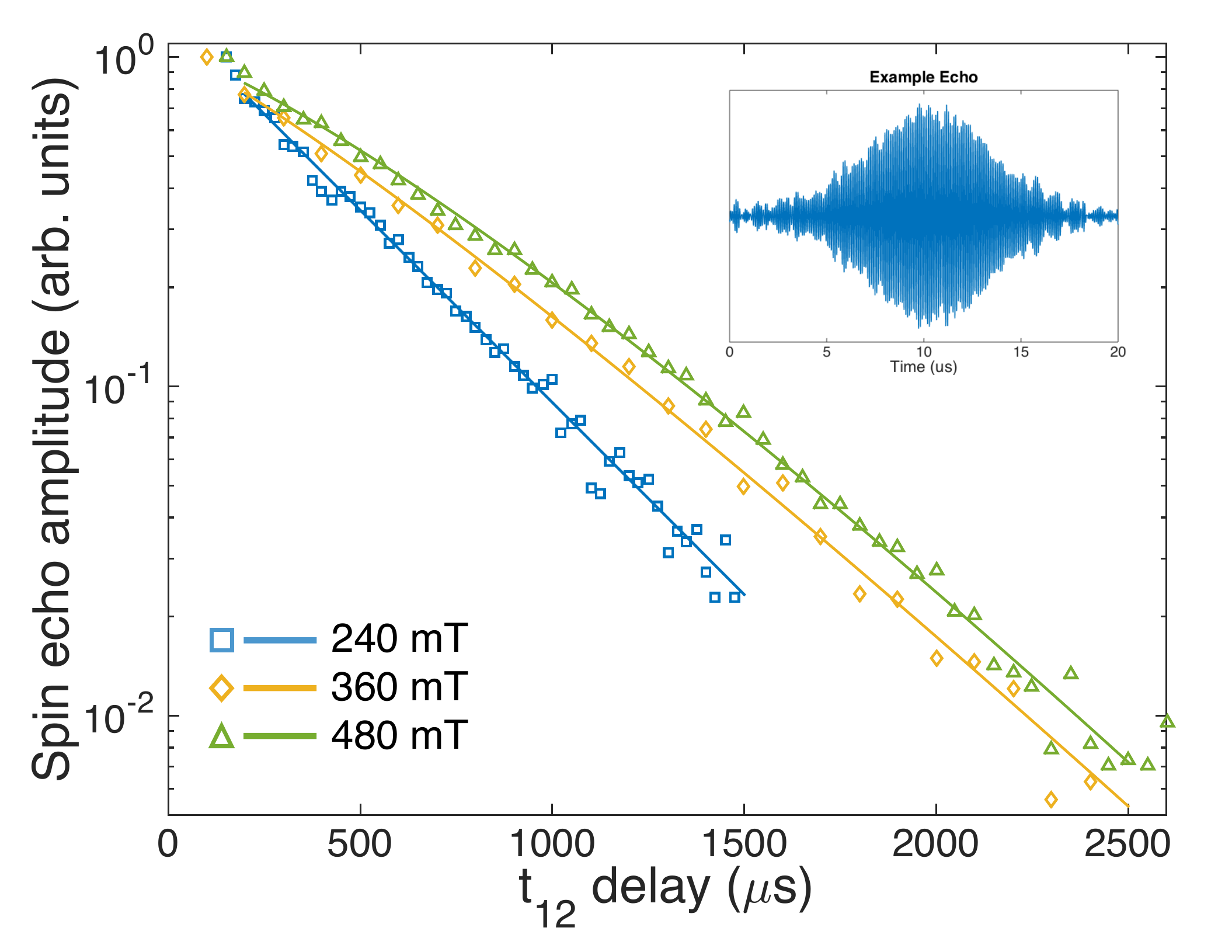}
\caption{Typical nuclear spin echo decays for increasing magnetic fields applied $20^\circ$ from the \textit{c}-axis. Here, the entire sequence is performed using all-optical manipulation of the spins.}
\label{fig:spinT2_alloptical}
\end{figure}

\fig{spinT2_alloptical} shows representative all-optical Raman echo decays for increasing applied magnetic fields applied $20^\circ$ from the \textit{c}-axis. Fitting to a Mims decay to describe the observed non-exponential behavior gave spin coherence times up to 1 ms at 480 mT, which was the maximum field available for this experiment. We note that moving the field off axis increases the magnetic sensitivity of the transition and therefore we expect a decrease of the spin coherence time in this regime. This configuration also reduces the ground state splitting, which would increase the contribution from Yb-Yb electron spin interactions. We expect to extend this coherence time with increasing magnetic fields as observed in the optical coherence measurements.

\section{Summary and Conclusion}
\label{sec:conclusion}

In this work, we have presented measurements assessing \ybyvo for use in quantum interfaces focusing on the strength of the optical transitions, the energy level structure, and the coherence properties of the optical and spin transitions.

From optical absorption measurements, we extract oscillator strengths in the upper range of those observed in REIs and larger than those observed for \ybion doped into \yso and YAG \cite{Welinski2016a,Bottger2016a}. This oscillator strength is promising for detecting and manipulating single ions coupled to nanophotonic cavities. The combination of large oscillator strengths and narrow inhomogeneous broadening in this material gives rise to exceptionally large absorption coefficients for this material. Significantly, the peak absorption of 450 cm$^{-1}$ is within a factor of 2 of the absorption in recently studied stoichiometric rare-earth crystals \cite{Ahlefeldt2016}, even though the ion concentration is a factor of $10^4$ more dilute. This absorption coefficient is promising for ensemble-based memories in bulk samples and reaching the impedance-matched regime necessary to achieve high-efficiency nanophotonic quantum memories. We observe a large branching ratio for relaxation directly to the ground state for the $^2F_{5/2}(0) \rightarrow {}^{2}F_{7/2}(0)$ transition, which is appealing in the context of Purcell enhancement of the emission rate in a nanocavity \cite{TZhong2015,TZhong2017}.

The large hyperfine couplings and narrow inhomogeneous linewidths give rise to resolved optical-hyperfine transitions. This is useful for addressing and manipulating single transitions and states without additional preparation. The large nuclear spin transition splittings are also useful in the context of off-resonance memory schemes and high-bandwidth spin wave storage. We completed a characterization of the energy level structure by determining the excited-state spin Hamiltonian. The knowledge of the full spin Hamiltonian is essential for designing optimal quantum interfaces because it facilitates the prediction and engineering of lambda systems or highly cyclic transitions. The symmetry of the crystal gives rise to the strong selection rules on the optical transitions. These selection rules are advantageous for engineering cyclic transitions \cite{Bartholomew2016}, especially in the context of a nanocavity that can preferentially enhance emission along one polarization. 

For protocols requiring additional nuclear spin states, the ${}^{173}\mathrm{Yb}$ isotope could be of interest because it has a nuclear spin of 5/2. The knowledge of the spin hamiltonian for the \ybion isotope allows the spin Hamiltonian for the $\mathrm{{}^{173}Yb}$ isotope to be approximated by scaling the hyperfine splittings by the ratio of the nuclear magnetic moments of the isotopes \cite{Elliott1953}. 

The optical coherence times are already sufficient for use in quantum applications and are promising from the perspective of reaching transform-limited photons in a nanocavity setting. The observed spin coherence properties show potential for use in spin-wave quantum memories and for single rare-earth ion qubits. In the linear Zeeman regime, the optical coherence times are limited by the superhyperfine interaction in the material. While a higher magnetic field regime was not available for the experimental configuration at the time of measurement, we expect the spin coherence times to increase with applied magnetic field as we completely freeze out the electron spin contribution. Further extension of the nuclear spin coherence time is predicted through the use of dynamic decoupling techniques \cite{MZhong2015}.  We note that the doping density used in these measurements (100 ppm) is relatively high  compared to many of the materials used for REI quantum technologies. We expect that by going to lower doping density samples we will reduce the contribution from Yb spin-spin interactions and observe longer optical and spin coherence times in the low field regime. Furthermore, by moving to lower temperatures, we can hope to freeze out the electron spins at lower magnetic fields. While this work focused on the optical and nuclear spin coherence properties in the linear Zeeman regime, we calculate that a zero-field ZEFOZ transition exists for one of the ground state transitions, which is expected to lead to an enhancement of coherence \cite{Ortu2017,Rakonjac2018a}. This configuration would also have the advantage of a strong electron spin transition between the ground states. Future studies are warranted to investigate the spin coherence properties of this material at lower doping densities, temperatures, and different magnetic field regimes. 

In summary, we find that \ybyvo is a promising material for REI-based quantum interfaces, such as ensemble-based quantum memories, microwave-to-optical transduction, and single REIs in nanophotonic cavities.

\section{Acknowledgements}

This work was funded by Office of Naval Research Young investigator Award N00014-16- 1-2676, Northrop Grumman, Montana Research and Economic Development Initiative, and the National Science Foundation under grants PHY-1415628 and CHE-1416454. J.G.B. acknowledges the support of the American Australian Association's Northrop Grumman Fellowship. J.M.K. and J.G.B would like to thank R.L.C., C.W.T., and P.J.T.W. for their hospitality throughout the trip to Bozeman during which part of this worked was performed. We would also like to acknowledge the expertise of Dr.~Yunbin Guan who was essential support for the SIMS analysis of the samples, and Dr.~Paul Oyala who assisted with preliminary bulk EPR measurements of this material.

\bibliographystyle{apsrev4-nourl}
\bibliography{/Users/Kindem/Documents/_Caltech/MendeleyBib/library}

\begin{thebibliography}{55}%
\makeatletter
\providecommand \@ifxundefined [1]{%
 \@ifx{#1\undefined}
}%
\providecommand \@ifnum [1]{%
 \ifnum #1\expandafter \@firstoftwo
 \else \expandafter \@secondoftwo
 \fi
}%
\providecommand \@ifx [1]{%
 \ifx #1\expandafter \@firstoftwo
 \else \expandafter \@secondoftwo
 \fi
}%
\providecommand \natexlab [1]{#1}%
\providecommand \enquote  [1]{``#1''}%
\providecommand \bibnamefont  [1]{#1}%
\providecommand \bibfnamefont [1]{#1}%
\providecommand \citenamefont [1]{#1}%
\providecommand \href@noop [0]{\@secondoftwo}%
\providecommand \href [0]{\begingroup \@sanitize@url \@href}%
\providecommand \@href[1]{\@@startlink{#1}\@@href}%
\providecommand \@@href[1]{\endgroup#1\@@endlink}%
\providecommand \@sanitize@url [0]{\catcode `\\12\catcode `\$12\catcode
  `\&12\catcode `\#12\catcode `\^12\catcode `\_12\catcode `\%12\relax}%
\providecommand \@@startlink[1]{}%
\providecommand \@@endlink[0]{}%
\providecommand \url  [0]{\begingroup\@sanitize@url \@url }%
\providecommand \@url [1]{\endgroup\@href {#1}{\urlprefix }}%
\providecommand \urlprefix  [0]{URL }%
\providecommand \Eprint [0]{\href }%
\providecommand \doibase [0]{http://dx.doi.org/}%
\providecommand \selectlanguage [0]{\@gobble}%
\providecommand \bibinfo  [0]{\@secondoftwo}%
\providecommand \bibfield  [0]{\@secondoftwo}%
\providecommand \translation [1]{[#1]}%
\providecommand \BibitemOpen [0]{}%
\providecommand \bibitemStop [0]{}%
\providecommand \bibitemNoStop [0]{.\EOS\space}%
\providecommand \EOS [0]{\spacefactor3000\relax}%
\providecommand \BibitemShut  [1]{\csname bibitem#1\endcsname}%
\let\auto@bib@innerbib\@empty
%</preamble>
\bibitem [{\citenamefont {Kimble}(2008)}]{Kimble2008}%
  \BibitemOpen
  \bibfield  {author} {\bibinfo {author} {\bibfnamefont {H.~J.}\ \bibnamefont
  {Kimble}},\ }\href {\doibase 10.1038/nature07127} {\bibfield  {journal}
  {\bibinfo  {journal} {Nature}\ }\textbf {\bibinfo {volume} {453}},\ \bibinfo
  {pages} {1023} (\bibinfo {year} {2008})}\BibitemShut {NoStop}%
\bibitem [{\citenamefont {Northup}\ and\ \citenamefont
  {Blatt}(2014)}]{Northup2014}%
  \BibitemOpen
  \bibfield  {author} {\bibinfo {author} {\bibfnamefont {T.~E.}\ \bibnamefont
  {Northup}}\ and\ \bibinfo {author} {\bibfnamefont {R.}~\bibnamefont
  {Blatt}},\ }\href {\doibase 10.1038/nphoton.2014.53} {\bibfield  {journal}
  {\bibinfo  {journal} {Nature Photonics}\ }\textbf {\bibinfo {volume} {8}},\
  \bibinfo {pages} {356} (\bibinfo {year} {2014})}\BibitemShut {NoStop}%
\bibitem [{\citenamefont {O'Brien}\ \emph {et~al.}(2009)\citenamefont
  {O'Brien}, \citenamefont {Furusawa},\ and\ \citenamefont
  {Vu{\v{c}}kovi{\'{c}}}}]{OBrien2009}%
  \BibitemOpen
  \bibfield  {author} {\bibinfo {author} {\bibfnamefont {J.~L.}\ \bibnamefont
  {O'Brien}}, \bibinfo {author} {\bibfnamefont {A.}~\bibnamefont {Furusawa}}, \
  and\ \bibinfo {author} {\bibfnamefont {J.}~\bibnamefont
  {Vu{\v{c}}kovi{\'{c}}}},\ }\href {\doibase 10.1038/nphoton.2009.229}
  {\bibfield  {journal} {\bibinfo  {journal} {Nature Photonics}\ }\textbf
  {\bibinfo {volume} {3}},\ \bibinfo {pages} {687} (\bibinfo {year}
  {2009})}\BibitemShut {NoStop}%
\bibitem [{\citenamefont {Andrews}\ \emph {et~al.}(2014)\citenamefont
  {Andrews}, \citenamefont {Peterson}, \citenamefont {Purdy}, \citenamefont
  {Cicak}, \citenamefont {Simmonds}, \citenamefont {Regal},\ and\ \citenamefont
  {Lehnert}}]{Andrews2014}%
  \BibitemOpen
  \bibfield  {author} {\bibinfo {author} {\bibfnamefont {R.~W.}\ \bibnamefont
  {Andrews}}, \bibinfo {author} {\bibfnamefont {R.~W.}\ \bibnamefont
  {Peterson}}, \bibinfo {author} {\bibfnamefont {T.~P.}\ \bibnamefont {Purdy}},
  \bibinfo {author} {\bibfnamefont {K.}~\bibnamefont {Cicak}}, \bibinfo
  {author} {\bibfnamefont {R.~W.}\ \bibnamefont {Simmonds}}, \bibinfo {author}
  {\bibfnamefont {C.~A.}\ \bibnamefont {Regal}}, \ and\ \bibinfo {author}
  {\bibfnamefont {K.~W.}\ \bibnamefont {Lehnert}},\ }\href {\doibase
  10.1038/nphys2911} {\bibfield  {journal} {\bibinfo  {journal} {Nature
  Physics}\ }\textbf {\bibinfo {volume} {10}},\ \bibinfo {pages} {321}
  (\bibinfo {year} {2014})}\BibitemShut {NoStop}%
\bibitem [{\citenamefont {Equall}\ \emph {et~al.}(1994)\citenamefont {Equall},
  \citenamefont {Sun}, \citenamefont {Cone},\ and\ \citenamefont
  {MacFarlane}}]{Equall1994}%
  \BibitemOpen
  \bibfield  {author} {\bibinfo {author} {\bibfnamefont {R.~W.}\ \bibnamefont
  {Equall}}, \bibinfo {author} {\bibfnamefont {Y.}~\bibnamefont {Sun}},
  \bibinfo {author} {\bibfnamefont {R.~L.}\ \bibnamefont {Cone}}, \ and\
  \bibinfo {author} {\bibfnamefont {R.~M.}\ \bibnamefont {MacFarlane}},\ }\href
  {\doibase 10.1103/PhysRevLett.72.2179} {\bibfield  {journal} {\bibinfo
  {journal} {Physical Review Letters}\ }\textbf {\bibinfo {volume} {72}},\
  \bibinfo {pages} {2179} (\bibinfo {year} {1994})}\BibitemShut {NoStop}%
\bibitem [{\citenamefont {B{\"{o}}ttger}\ \emph {et~al.}(2009)\citenamefont
  {B{\"{o}}ttger}, \citenamefont {Thiel}, \citenamefont {Cone},\ and\
  \citenamefont {Sun}}]{Bottger2009a}%
  \BibitemOpen
  \bibfield  {author} {\bibinfo {author} {\bibfnamefont {T.}~\bibnamefont
  {B{\"{o}}ttger}}, \bibinfo {author} {\bibfnamefont {C.~W.}\ \bibnamefont
  {Thiel}}, \bibinfo {author} {\bibfnamefont {R.~L.}\ \bibnamefont {Cone}}, \
  and\ \bibinfo {author} {\bibfnamefont {Y.}~\bibnamefont {Sun}},\ }\href
  {\doibase 10.1103/PhysRevB.79.115104} {\bibfield  {journal} {\bibinfo
  {journal} {Physical Review B}\ }\textbf {\bibinfo {volume} {79}},\ \bibinfo
  {pages} {115104} (\bibinfo {year} {2009})}\BibitemShut {NoStop}%
\bibitem [{\citenamefont {Zhong}\ \emph
  {et~al.}(2015{\natexlab{a}})\citenamefont {Zhong}, \citenamefont {Hedges},
  \citenamefont {Ahlefeldt}, \citenamefont {Bartholomew}, \citenamefont
  {Beavan}, \citenamefont {Wittig}, \citenamefont {Longdell},\ and\
  \citenamefont {Sellars}}]{MZhong2015}%
  \BibitemOpen
  \bibfield  {author} {\bibinfo {author} {\bibfnamefont {M.}~\bibnamefont
  {Zhong}}, \bibinfo {author} {\bibfnamefont {M.~P.}\ \bibnamefont {Hedges}},
  \bibinfo {author} {\bibfnamefont {R.~L.}\ \bibnamefont {Ahlefeldt}}, \bibinfo
  {author} {\bibfnamefont {J.~G.}\ \bibnamefont {Bartholomew}}, \bibinfo
  {author} {\bibfnamefont {S.~E.}\ \bibnamefont {Beavan}}, \bibinfo {author}
  {\bibfnamefont {S.~M.}\ \bibnamefont {Wittig}}, \bibinfo {author}
  {\bibfnamefont {J.~J.}\ \bibnamefont {Longdell}}, \ and\ \bibinfo {author}
  {\bibfnamefont {M.~J.}\ \bibnamefont {Sellars}},\ }\href {\doibase
  10.1038/nature14025} {\bibfield  {journal} {\bibinfo  {journal} {Nature}\
  }\textbf {\bibinfo {volume} {517}},\ \bibinfo {pages} {177} (\bibinfo {year}
  {2015}{\natexlab{a}})}\BibitemShut {NoStop}%
\bibitem [{\citenamefont {Ran{\v{c}}i{\'{c}}}\ \emph
  {et~al.}(2017)\citenamefont {Ran{\v{c}}i{\'{c}}}, \citenamefont {Hedges},
  \citenamefont {Ahlefeldt},\ and\ \citenamefont {Sellars}}]{Rancic2017}%
  \BibitemOpen
  \bibfield  {author} {\bibinfo {author} {\bibfnamefont {M.}~\bibnamefont
  {Ran{\v{c}}i{\'{c}}}}, \bibinfo {author} {\bibfnamefont {M.~P.}\ \bibnamefont
  {Hedges}}, \bibinfo {author} {\bibfnamefont {R.~L.}\ \bibnamefont
  {Ahlefeldt}}, \ and\ \bibinfo {author} {\bibfnamefont {M.~J.}\ \bibnamefont
  {Sellars}},\ }\href {\doibase 10.1038/nphys4254} {\bibfield  {journal}
  {\bibinfo  {journal} {Nature Physics}\ }\textbf {\bibinfo {volume} {14}}
  (\bibinfo {year} {2017})}\BibitemShut {NoStop}%
\bibitem [{\citenamefont {Tittel}\ \emph {et~al.}(2009)\citenamefont {Tittel},
  \citenamefont {Afzelius}, \citenamefont {Chaneli{\'{e}}re}, \citenamefont
  {Cone}, \citenamefont {Kr{\"{o}}ll}, \citenamefont {Moiseev},\ and\
  \citenamefont {Sellars}}]{Tittel2009}%
  \BibitemOpen
  \bibfield  {author} {\bibinfo {author} {\bibfnamefont {W.}~\bibnamefont
  {Tittel}}, \bibinfo {author} {\bibfnamefont {M.}~\bibnamefont {Afzelius}},
  \bibinfo {author} {\bibfnamefont {T.}~\bibnamefont {Chaneli{\'{e}}re}},
  \bibinfo {author} {\bibfnamefont {R.}~\bibnamefont {Cone}}, \bibinfo {author}
  {\bibfnamefont {S.}~\bibnamefont {Kr{\"{o}}ll}}, \bibinfo {author}
  {\bibfnamefont {S.}~\bibnamefont {Moiseev}}, \ and\ \bibinfo {author}
  {\bibfnamefont {M.}~\bibnamefont {Sellars}},\ }\href {\doibase
  10.1002/lpor.200810056} {\bibfield  {journal} {\bibinfo  {journal} {Laser
  {\&} Photonics Reviews}\ }\textbf {\bibinfo {volume} {4}},\ \bibinfo {pages}
  {244} (\bibinfo {year} {2009})}\BibitemShut {NoStop}%
\bibitem [{\citenamefont {Hedges}\ \emph {et~al.}(2010)\citenamefont {Hedges},
  \citenamefont {Longdell}, \citenamefont {Li},\ and\ \citenamefont
  {Sellars}}]{Hedges2010}%
  \BibitemOpen
  \bibfield  {author} {\bibinfo {author} {\bibfnamefont {M.~P.}\ \bibnamefont
  {Hedges}}, \bibinfo {author} {\bibfnamefont {J.~J.}\ \bibnamefont
  {Longdell}}, \bibinfo {author} {\bibfnamefont {Y.}~\bibnamefont {Li}}, \ and\
  \bibinfo {author} {\bibfnamefont {M.~J.}\ \bibnamefont {Sellars}},\ }\href
  {\doibase 10.1038/nature09081} {\bibfield  {journal} {\bibinfo  {journal}
  {Nature}\ }\textbf {\bibinfo {volume} {465}},\ \bibinfo {pages} {1052}
  (\bibinfo {year} {2010})}\BibitemShut {NoStop}%
\bibitem [{\citenamefont {Sabooni}\ \emph {et~al.}(2013)\citenamefont
  {Sabooni}, \citenamefont {Li}, \citenamefont {Kr{\"{o}}ll},\ and\
  \citenamefont {Rippe}}]{Sabooni2013}%
  \BibitemOpen
  \bibfield  {author} {\bibinfo {author} {\bibfnamefont {M.}~\bibnamefont
  {Sabooni}}, \bibinfo {author} {\bibfnamefont {Q.}~\bibnamefont {Li}},
  \bibinfo {author} {\bibfnamefont {S.}~\bibnamefont {Kr{\"{o}}ll}}, \ and\
  \bibinfo {author} {\bibfnamefont {L.}~\bibnamefont {Rippe}},\ }\href
  {\doibase 10.1103/PhysRevLett.110.133604} {\bibfield  {journal} {\bibinfo
  {journal} {Physical Review Letters}\ }\textbf {\bibinfo {volume} {110}},\
  \bibinfo {pages} {133604} (\bibinfo {year} {2013})}\BibitemShut {NoStop}%
\bibitem [{\citenamefont {Saglamyurek}\ \emph {et~al.}(2014)\citenamefont
  {Saglamyurek}, \citenamefont {Sinclair}, \citenamefont {Slater},
  \citenamefont {Heshami}, \citenamefont {Oblak},\ and\ \citenamefont
  {Tittel}}]{Saglamyurek2014}%
  \BibitemOpen
  \bibfield  {author} {\bibinfo {author} {\bibfnamefont {E.}~\bibnamefont
  {Saglamyurek}}, \bibinfo {author} {\bibfnamefont {N.}~\bibnamefont
  {Sinclair}}, \bibinfo {author} {\bibfnamefont {J.~A.}\ \bibnamefont
  {Slater}}, \bibinfo {author} {\bibfnamefont {K.}~\bibnamefont {Heshami}},
  \bibinfo {author} {\bibfnamefont {D.}~\bibnamefont {Oblak}}, \ and\ \bibinfo
  {author} {\bibfnamefont {W.}~\bibnamefont {Tittel}},\ }\href {\doibase
  10.1088/1367-2630/16/6/065019} {\bibfield  {journal} {\bibinfo  {journal}
  {New Journal of Physics}\ }\textbf {\bibinfo {volume} {16}} (\bibinfo {year}
  {2014})}\BibitemShut {NoStop}%
\bibitem [{\citenamefont {G{\"{u}}ndogan}\ \emph {et~al.}(2015)\citenamefont
  {G{\"{u}}ndogan}, \citenamefont {Ledingham}, \citenamefont {Kutluer},
  \citenamefont {Mazzera},\ and\ \citenamefont {de~Riedmatten}}]{Gundogan2015}%
  \BibitemOpen
  \bibfield  {author} {\bibinfo {author} {\bibfnamefont {M.}~\bibnamefont
  {G{\"{u}}ndogan}}, \bibinfo {author} {\bibfnamefont {P.~M.}\ \bibnamefont
  {Ledingham}}, \bibinfo {author} {\bibfnamefont {K.}~\bibnamefont {Kutluer}},
  \bibinfo {author} {\bibfnamefont {M.}~\bibnamefont {Mazzera}}, \ and\
  \bibinfo {author} {\bibfnamefont {H.}~\bibnamefont {de~Riedmatten}},\ }\href
  {\doibase 10.1103/PhysRevLett.114.230501} {\bibfield  {journal} {\bibinfo
  {journal} {Physical Review Letters}\ }\textbf {\bibinfo {volume} {114}},\
  \bibinfo {pages} {230501} (\bibinfo {year} {2015})}\BibitemShut {NoStop}%
\bibitem [{\citenamefont {Laplane}\ \emph {et~al.}(2016)\citenamefont
  {Laplane}, \citenamefont {Jobez}, \citenamefont {Etesse}, \citenamefont
  {Timoney}, \citenamefont {Gisin},\ and\ \citenamefont
  {Afzelius}}]{Laplane2016}%
  \BibitemOpen
  \bibfield  {author} {\bibinfo {author} {\bibfnamefont {C.}~\bibnamefont
  {Laplane}}, \bibinfo {author} {\bibfnamefont {P.}~\bibnamefont {Jobez}},
  \bibinfo {author} {\bibfnamefont {J.}~\bibnamefont {Etesse}}, \bibinfo
  {author} {\bibfnamefont {N.}~\bibnamefont {Timoney}}, \bibinfo {author}
  {\bibfnamefont {N.}~\bibnamefont {Gisin}}, \ and\ \bibinfo {author}
  {\bibfnamefont {M.}~\bibnamefont {Afzelius}},\ }\href {\doibase
  10.1088/1367-2630/18/1/013006} {\bibfield  {journal} {\bibinfo  {journal}
  {New Journal of Physics}\ }\textbf {\bibinfo {volume} {18}},\ \bibinfo
  {pages} {013006} (\bibinfo {year} {2016})}\BibitemShut {NoStop}%
\bibitem [{\citenamefont {Lauritzen}\ \emph {et~al.}(2008)\citenamefont
  {Lauritzen}, \citenamefont {Hastings-Simon}, \citenamefont {de~Riedmatten},
  \citenamefont {Afzelius},\ and\ \citenamefont {Gisin}}]{Lauritzen2008}%
  \BibitemOpen
  \bibfield  {author} {\bibinfo {author} {\bibfnamefont {B.}~\bibnamefont
  {Lauritzen}}, \bibinfo {author} {\bibfnamefont {S.~R.}\ \bibnamefont
  {Hastings-Simon}}, \bibinfo {author} {\bibfnamefont {H.}~\bibnamefont
  {de~Riedmatten}}, \bibinfo {author} {\bibfnamefont {M.}~\bibnamefont
  {Afzelius}}, \ and\ \bibinfo {author} {\bibfnamefont {N.}~\bibnamefont
  {Gisin}},\ }\href {\doibase 10.1103/PhysRevA.78.043402} {\bibfield  {journal}
  {\bibinfo  {journal} {Physical Review A}\ }\textbf {\bibinfo {volume} {78}},\
  \bibinfo {pages} {043402} (\bibinfo {year} {2008})}\BibitemShut {NoStop}%
\bibitem [{\citenamefont {Baldit}\ \emph {et~al.}(2010)\citenamefont {Baldit},
  \citenamefont {Bencheikh}, \citenamefont {Monnier}, \citenamefont
  {Briaudeau}, \citenamefont {Levenson}, \citenamefont {Crozatier},
  \citenamefont {Lorger{\'{e}}}, \citenamefont {Bretenaker}, \citenamefont {{Le
  Gou{\"{e}}t}}, \citenamefont {Guillot-No{\"{e}}l},\ and\ \citenamefont
  {Goldner}}]{Baldit2010}%
  \BibitemOpen
  \bibfield  {author} {\bibinfo {author} {\bibfnamefont {E.}~\bibnamefont
  {Baldit}}, \bibinfo {author} {\bibfnamefont {K.}~\bibnamefont {Bencheikh}},
  \bibinfo {author} {\bibfnamefont {P.}~\bibnamefont {Monnier}}, \bibinfo
  {author} {\bibfnamefont {S.}~\bibnamefont {Briaudeau}}, \bibinfo {author}
  {\bibfnamefont {J.~A.}\ \bibnamefont {Levenson}}, \bibinfo {author}
  {\bibfnamefont {V.}~\bibnamefont {Crozatier}}, \bibinfo {author}
  {\bibfnamefont {I.}~\bibnamefont {Lorger{\'{e}}}}, \bibinfo {author}
  {\bibfnamefont {F.}~\bibnamefont {Bretenaker}}, \bibinfo {author}
  {\bibfnamefont {J.~L.}\ \bibnamefont {{Le Gou{\"{e}}t}}}, \bibinfo {author}
  {\bibfnamefont {O.}~\bibnamefont {Guillot-No{\"{e}}l}}, \ and\ \bibinfo
  {author} {\bibfnamefont {P.}~\bibnamefont {Goldner}},\ }\href {\doibase
  10.1103/PhysRevB.81.144303} {\bibfield  {journal} {\bibinfo  {journal}
  {Physical Review B}\ }\textbf {\bibinfo {volume} {81}},\ \bibinfo {pages}
  {144303} (\bibinfo {year} {2010})}\BibitemShut {NoStop}%
\bibitem [{\citenamefont {Welinski}\ \emph {et~al.}(2016)\citenamefont
  {Welinski}, \citenamefont {Ferrier}, \citenamefont {Afzelius},\ and\
  \citenamefont {Goldner}}]{Welinski2016a}%
  \BibitemOpen
  \bibfield  {author} {\bibinfo {author} {\bibfnamefont {S.}~\bibnamefont
  {Welinski}}, \bibinfo {author} {\bibfnamefont {A.}~\bibnamefont {Ferrier}},
  \bibinfo {author} {\bibfnamefont {M.}~\bibnamefont {Afzelius}}, \ and\
  \bibinfo {author} {\bibfnamefont {P.}~\bibnamefont {Goldner}},\ }\href
  {\doibase 10.1103/PhysRevB.94.155116} {\bibfield  {journal} {\bibinfo
  {journal} {Physical Review B}\ }\textbf {\bibinfo {volume} {94}},\ \bibinfo
  {pages} {155116} (\bibinfo {year} {2016})}\BibitemShut {NoStop}%
\bibitem [{\citenamefont {Tiranov}\ \emph {et~al.}(2017)\citenamefont
  {Tiranov}, \citenamefont {Ortu}, \citenamefont {Welinski}, \citenamefont
  {Ferrier}, \citenamefont {Goldner}, \citenamefont {Gisin},\ and\
  \citenamefont {Afzelius}}]{Tiranov2017}%
  \BibitemOpen
  \bibfield  {author} {\bibinfo {author} {\bibfnamefont {A.}~\bibnamefont
  {Tiranov}}, \bibinfo {author} {\bibfnamefont {A.}~\bibnamefont {Ortu}},
  \bibinfo {author} {\bibfnamefont {S.}~\bibnamefont {Welinski}}, \bibinfo
  {author} {\bibfnamefont {A.}~\bibnamefont {Ferrier}}, \bibinfo {author}
  {\bibfnamefont {P.}~\bibnamefont {Goldner}}, \bibinfo {author} {\bibfnamefont
  {N.}~\bibnamefont {Gisin}}, \ and\ \bibinfo {author} {\bibfnamefont
  {M.}~\bibnamefont {Afzelius}},\ }\href {http://arxiv.org/abs/1712.08616}
  {}\Eprint {http://arxiv.org/abs/1712.08616} {arXiv:1712.08616}  (\bibinfo
  {year} {2017})\BibitemShut {NoStop}%
\bibitem [{\citenamefont {Lim}\ \emph {et~al.}(2018)\citenamefont {Lim},
  \citenamefont {Welinski}, \citenamefont {Ferrier}, \citenamefont {Goldner},\
  and\ \citenamefont {Morton}}]{Lim2018}%
  \BibitemOpen
  \bibfield  {author} {\bibinfo {author} {\bibfnamefont {H.-J.}\ \bibnamefont
  {Lim}}, \bibinfo {author} {\bibfnamefont {S.}~\bibnamefont {Welinski}},
  \bibinfo {author} {\bibfnamefont {A.}~\bibnamefont {Ferrier}}, \bibinfo
  {author} {\bibfnamefont {P.}~\bibnamefont {Goldner}}, \ and\ \bibinfo
  {author} {\bibfnamefont {J.~J.}\ \bibnamefont {Morton}},\ }\href {\doibase
  10.1103/PhysRevB.97.064409} {\bibfield  {journal} {\bibinfo  {journal}
  {Physical Review B}\ }\textbf {\bibinfo {volume} {97}},\ \bibinfo {pages}
  {064409} (\bibinfo {year} {2018})}\BibitemShut {NoStop}%
\bibitem [{\citenamefont {Ortu}\ \emph {et~al.}(2017)\citenamefont {Ortu},
  \citenamefont {Tiranov}, \citenamefont {Welinski}, \citenamefont
  {Fr{\"{o}}wis}, \citenamefont {Gisin}, \citenamefont {Ferrier}, \citenamefont
  {Goldner},\ and\ \citenamefont {Afzelius}}]{Ortu2017}%
  \BibitemOpen
  \bibfield  {author} {\bibinfo {author} {\bibfnamefont {A.}~\bibnamefont
  {Ortu}}, \bibinfo {author} {\bibfnamefont {A.}~\bibnamefont {Tiranov}},
  \bibinfo {author} {\bibfnamefont {S.}~\bibnamefont {Welinski}}, \bibinfo
  {author} {\bibfnamefont {F.}~\bibnamefont {Fr{\"{o}}wis}}, \bibinfo {author}
  {\bibfnamefont {N.}~\bibnamefont {Gisin}}, \bibinfo {author} {\bibfnamefont
  {A.}~\bibnamefont {Ferrier}}, \bibinfo {author} {\bibfnamefont
  {P.}~\bibnamefont {Goldner}}, \ and\ \bibinfo {author} {\bibfnamefont
  {M.}~\bibnamefont {Afzelius}},\ }\href {http://arxiv.org/abs/1712.08615}
  {}\Eprint {http://arxiv.org/abs/1712.08615} {arXiv:1712.08615}  (\bibinfo
  {year} {2017})\BibitemShut {NoStop}%
\bibitem [{\citenamefont {Kis}\ \emph {et~al.}(2014)\citenamefont {Kis},
  \citenamefont {Mandula}, \citenamefont {Lengyel}, \citenamefont {Hajdara},
  \citenamefont {Kovacs},\ and\ \citenamefont {Imlau}}]{Kis2014}%
  \BibitemOpen
  \bibfield  {author} {\bibinfo {author} {\bibfnamefont {Z.}~\bibnamefont
  {Kis}}, \bibinfo {author} {\bibfnamefont {G.}~\bibnamefont {Mandula}},
  \bibinfo {author} {\bibfnamefont {K.}~\bibnamefont {Lengyel}}, \bibinfo
  {author} {\bibfnamefont {I.}~\bibnamefont {Hajdara}}, \bibinfo {author}
  {\bibfnamefont {L.}~\bibnamefont {Kovacs}}, \ and\ \bibinfo {author}
  {\bibfnamefont {M.}~\bibnamefont {Imlau}},\ }\href {\doibase
  10.1016/j.optmat.2014.09.022} {\bibfield  {journal} {\bibinfo  {journal}
  {Optical Materials}\ }\textbf {\bibinfo {volume} {37}},\ \bibinfo {pages}
  {845} (\bibinfo {year} {2014})}\BibitemShut {NoStop}%
\bibitem [{\citenamefont {B{\"{o}}ttger}\ \emph {et~al.}(2016)\citenamefont
  {B{\"{o}}ttger}, \citenamefont {Thiel}, \citenamefont {Cone}, \citenamefont
  {Sun},\ and\ \citenamefont {Faraon}}]{Bottger2016a}%
  \BibitemOpen
  \bibfield  {author} {\bibinfo {author} {\bibfnamefont {T.}~\bibnamefont
  {B{\"{o}}ttger}}, \bibinfo {author} {\bibfnamefont {C.~W.}\ \bibnamefont
  {Thiel}}, \bibinfo {author} {\bibfnamefont {R.~L.}\ \bibnamefont {Cone}},
  \bibinfo {author} {\bibfnamefont {Y.}~\bibnamefont {Sun}}, \ and\ \bibinfo
  {author} {\bibfnamefont {A.}~\bibnamefont {Faraon}},\ }\href {\doibase
  10.1103/PhysRevB.94.045134} {\bibfield  {journal} {\bibinfo  {journal}
  {Physical Review B}\ }\textbf {\bibinfo {volume} {94}},\ \bibinfo {pages}
  {045134} (\bibinfo {year} {2016})}\BibitemShut {NoStop}%
\bibitem [{\citenamefont {de~Riedmatten}\ \emph {et~al.}(2008)\citenamefont
  {de~Riedmatten}, \citenamefont {Afzelius}, \citenamefont {Staudt},
  \citenamefont {Simon},\ and\ \citenamefont {Gisin}}]{DeRiedmatten2008}%
  \BibitemOpen
  \bibfield  {author} {\bibinfo {author} {\bibfnamefont {H.}~\bibnamefont
  {de~Riedmatten}}, \bibinfo {author} {\bibfnamefont {M.}~\bibnamefont
  {Afzelius}}, \bibinfo {author} {\bibfnamefont {M.~U.}\ \bibnamefont
  {Staudt}}, \bibinfo {author} {\bibfnamefont {C.}~\bibnamefont {Simon}}, \
  and\ \bibinfo {author} {\bibfnamefont {N.}~\bibnamefont {Gisin}},\ }\href
  {\doibase 10.1038/nature07607} {\bibfield  {journal} {\bibinfo  {journal}
  {Nature}\ }\textbf {\bibinfo {volume} {456}},\ \bibinfo {pages} {773}
  (\bibinfo {year} {2008})}\BibitemShut {NoStop}%
\bibitem [{\citenamefont {Zhong}\ \emph {et~al.}(2017)\citenamefont {Zhong},
  \citenamefont {Kindem}, \citenamefont {Bartholomew}, \citenamefont {Rochman},
  \citenamefont {Craiciu}, \citenamefont {Miyazono}, \citenamefont
  {Bettinelli}, \citenamefont {Cavalli}, \citenamefont {Verma}, \citenamefont
  {Nam}, \citenamefont {Marsili}, \citenamefont {Shaw}, \citenamefont {Beyer},\
  and\ \citenamefont {Faraon}}]{TZhong2017}%
  \BibitemOpen
  \bibfield  {author} {\bibinfo {author} {\bibfnamefont {T.}~\bibnamefont
  {Zhong}}, \bibinfo {author} {\bibfnamefont {J.~M.}\ \bibnamefont {Kindem}},
  \bibinfo {author} {\bibfnamefont {J.~G.}\ \bibnamefont {Bartholomew}},
  \bibinfo {author} {\bibfnamefont {J.}~\bibnamefont {Rochman}}, \bibinfo
  {author} {\bibfnamefont {I.}~\bibnamefont {Craiciu}}, \bibinfo {author}
  {\bibfnamefont {E.}~\bibnamefont {Miyazono}}, \bibinfo {author}
  {\bibfnamefont {M.}~\bibnamefont {Bettinelli}}, \bibinfo {author}
  {\bibfnamefont {E.}~\bibnamefont {Cavalli}}, \bibinfo {author} {\bibfnamefont
  {V.}~\bibnamefont {Verma}}, \bibinfo {author} {\bibfnamefont {S.~W.}\
  \bibnamefont {Nam}}, \bibinfo {author} {\bibfnamefont {F.}~\bibnamefont
  {Marsili}}, \bibinfo {author} {\bibfnamefont {M.~D.}\ \bibnamefont {Shaw}},
  \bibinfo {author} {\bibfnamefont {A.~D.}\ \bibnamefont {Beyer}}, \ and\
  \bibinfo {author} {\bibfnamefont {A.}~\bibnamefont {Faraon}},\ }\href@noop {}
  {\bibfield  {journal} {\bibinfo  {journal} {Science}\ }\textbf {\bibinfo
  {volume} {1395}},\ \bibinfo {pages} {1392} (\bibinfo {year}
  {2017})}\BibitemShut {NoStop}%
\bibitem [{\citenamefont {Zhong}\ \emph {et~al.}(2016)\citenamefont {Zhong},
  \citenamefont {Rochman}, \citenamefont {Kindem}, \citenamefont {Miyazono},\
  and\ \citenamefont {Faraon}}]{TZhong2016}%
  \BibitemOpen
  \bibfield  {author} {\bibinfo {author} {\bibfnamefont {T.}~\bibnamefont
  {Zhong}}, \bibinfo {author} {\bibfnamefont {J.}~\bibnamefont {Rochman}},
  \bibinfo {author} {\bibfnamefont {J.~M.}\ \bibnamefont {Kindem}}, \bibinfo
  {author} {\bibfnamefont {E.}~\bibnamefont {Miyazono}}, \ and\ \bibinfo
  {author} {\bibfnamefont {A.}~\bibnamefont {Faraon}},\ }\href {\doibase
  10.1364/OE.24.000536} {\bibfield  {journal} {\bibinfo  {journal} {Optics
  Express}\ }\textbf {\bibinfo {volume} {24}},\ \bibinfo {pages} {536}
  (\bibinfo {year} {2016})}\BibitemShut {NoStop}%
\bibitem [{\citenamefont {Kr{\"{a}}nkel}\ \emph {et~al.}(2004)\citenamefont
  {Kr{\"{a}}nkel}, \citenamefont {Fagundes-Peters}, \citenamefont {Fredrich},
  \citenamefont {Johannsen}, \citenamefont {Mond}, \citenamefont {Huber},
  \citenamefont {Bernhagen},\ and\ \citenamefont {Uecker}}]{Krankel2004}%
  \BibitemOpen
  \bibfield  {author} {\bibinfo {author} {\bibfnamefont {C.}~\bibnamefont
  {Kr{\"{a}}nkel}}, \bibinfo {author} {\bibfnamefont {D.}~\bibnamefont
  {Fagundes-Peters}}, \bibinfo {author} {\bibfnamefont {S.~T.}\ \bibnamefont
  {Fredrich}}, \bibinfo {author} {\bibfnamefont {J.}~\bibnamefont {Johannsen}},
  \bibinfo {author} {\bibfnamefont {M.}~\bibnamefont {Mond}}, \bibinfo {author}
  {\bibfnamefont {G.}~\bibnamefont {Huber}}, \bibinfo {author} {\bibfnamefont
  {M.}~\bibnamefont {Bernhagen}}, \ and\ \bibinfo {author} {\bibfnamefont
  {R.}~\bibnamefont {Uecker}},\ }\href {\doibase 10.1007/s00340-004-1635-y}
  {\bibfield  {journal} {\bibinfo  {journal} {Applied Physics B: Lasers and
  Optics}\ }\textbf {\bibinfo {volume} {79}},\ \bibinfo {pages} {543} (\bibinfo
  {year} {2004})}\BibitemShut {NoStop}%
\bibitem [{\citenamefont {Wyckoff}(1963)}]{Wyckoff1963}%
  \BibitemOpen
  \bibfield  {author} {\bibinfo {author} {\bibfnamefont {R.}~\bibnamefont
  {Wyckoff}},\ }\href@noop {} {\emph {\bibinfo {title} {{Crystal
  Structures}}}},\ \bibinfo {edition} {2nd}\ ed.\ (\bibinfo  {publisher} {John
  Wiley {\&} Sons},\ \bibinfo {address} {New York, NY},\ \bibinfo {year}
  {1963})\BibitemShut {NoStop}%
\bibitem [{\citenamefont {Pestryakov}\ \emph {et~al.}(2006)\citenamefont
  {Pestryakov}, \citenamefont {Petrov}, \citenamefont {Trunov}, \citenamefont
  {Kirpichnikov}, \citenamefont {Merzliakov}, \citenamefont {Laptev},\ and\
  \citenamefont {Matrosov}}]{Pestryakov2005}%
  \BibitemOpen
  \bibfield  {author} {\bibinfo {author} {\bibfnamefont {E.~V.}\ \bibnamefont
  {Pestryakov}}, \bibinfo {author} {\bibfnamefont {V.~V.}\ \bibnamefont
  {Petrov}}, \bibinfo {author} {\bibfnamefont {V.~I.}\ \bibnamefont {Trunov}},
  \bibinfo {author} {\bibfnamefont {A.~V.}\ \bibnamefont {Kirpichnikov}},
  \bibinfo {author} {\bibfnamefont {M.~A.}\ \bibnamefont {Merzliakov}},
  \bibinfo {author} {\bibfnamefont {A.~V.}\ \bibnamefont {Laptev}}, \ and\
  \bibinfo {author} {\bibfnamefont {V.~N.}\ \bibnamefont {Matrosov}},\ }\href
  {\doibase 10.1117/12.660705} {\bibfield  {journal} {\bibinfo  {journal}
  {Proc. of SPIE}\ }\textbf {\bibinfo {volume} {6054}} (\bibinfo {year}
  {2006})}\BibitemShut {NoStop}%
\bibitem [{\citenamefont {Abragam}\ and\ \citenamefont
  {Bleaney}(2012)}]{Abragam1970}%
  \BibitemOpen
  \bibfield  {author} {\bibinfo {author} {\bibfnamefont {A.}~\bibnamefont
  {Abragam}}\ and\ \bibinfo {author} {\bibfnamefont {B.}~\bibnamefont
  {Bleaney}},\ }\href {\doibase 10.1017/CBO9781107415324.004} {\emph {\bibinfo
  {title} {Electron Paramagnetic Resonance of Transition Ions}}}\ (\bibinfo
  {publisher} {Oxford University Press},\ \bibinfo {address} {Oxford},\
  \bibinfo {year} {2012})\BibitemShut {NoStop}%
\bibitem [{\citenamefont {Ranon}(1968)}]{Ranon1968}%
  \BibitemOpen
  \bibfield  {author} {\bibinfo {author} {\bibfnamefont {U.}~\bibnamefont
  {Ranon}},\ }\href {\doibase 10.1016/0375-9601(68)90218-1} {\bibfield
  {journal} {\bibinfo  {journal} {Physics Letters A}\ }\textbf {\bibinfo
  {volume} {28}},\ \bibinfo {pages} {228} (\bibinfo {year} {1968})}\BibitemShut
  {NoStop}%
\bibitem [{\citenamefont {Cook}\ \emph {et~al.}(2012)\citenamefont {Cook},
  \citenamefont {Martin}, \citenamefont {Brown-Heft}, \citenamefont {Garman},\
  and\ \citenamefont {Steck}}]{Cook2012}%
  \BibitemOpen
  \bibfield  {author} {\bibinfo {author} {\bibfnamefont {E.~C.}\ \bibnamefont
  {Cook}}, \bibinfo {author} {\bibfnamefont {P.~J.}\ \bibnamefont {Martin}},
  \bibinfo {author} {\bibfnamefont {T.~L.}\ \bibnamefont {Brown-Heft}},
  \bibinfo {author} {\bibfnamefont {J.~C.}\ \bibnamefont {Garman}}, \ and\
  \bibinfo {author} {\bibfnamefont {D.~A.}\ \bibnamefont {Steck}},\ }\href
  {\doibase 10.1063/1.3698003} {\bibfield  {journal} {\bibinfo  {journal}
  {Review of Scientific Instruments}\ }\textbf {\bibinfo {volume} {83}},\
  \bibinfo {pages} {043101} (\bibinfo {year} {2012})}\BibitemShut {NoStop}%
\bibitem [{\citenamefont {Liu}\ and\ \citenamefont {Jacquier}(2005)}]{Liu2005}%
  \BibitemOpen
  \bibinfo {editor} {\bibfnamefont {G.}~\bibnamefont {Liu}}\ and\ \bibinfo
  {editor} {\bibfnamefont {B.}~\bibnamefont {Jacquier}},\ eds.,\ \href
  {\doibase 10.1007/3-540-28209-2} {\emph {\bibinfo {title} {Spectroscopic
  Properties of Rare Earths in Optical Materials}}},\ \bibinfo {series}
  {Springer Series in Materials Science}, Vol.~\bibinfo {volume} {83}\
  (\bibinfo  {publisher} {Springer-Verlag},\ \bibinfo {address}
  {Berlin/Heidelberg},\ \bibinfo {year} {2005})\BibitemShut {NoStop}%
\bibitem [{\citenamefont {Wei}\ \emph {et~al.}(1996)\citenamefont {Wei},
  \citenamefont {Holmstrom}, \citenamefont {Manson}, \citenamefont {Martin},
  \citenamefont {He},\ and\ \citenamefont {Fisk}}]{Wei1996}%
  \BibitemOpen
  \bibfield  {author} {\bibinfo {author} {\bibfnamefont {C.}~\bibnamefont
  {Wei}}, \bibinfo {author} {\bibfnamefont {S.~A.}\ \bibnamefont {Holmstrom}},
  \bibinfo {author} {\bibfnamefont {N.~B.}\ \bibnamefont {Manson}}, \bibinfo
  {author} {\bibfnamefont {J.~P.~D.}\ \bibnamefont {Martin}}, \bibinfo {author}
  {\bibfnamefont {X.~F.}\ \bibnamefont {He}}, \ and\ \bibinfo {author}
  {\bibfnamefont {P.~T.~H.}\ \bibnamefont {Fisk}},\ }\href@noop {} {\bibfield
  {journal} {\bibinfo  {journal} {Appl. Magn. Reson.}\ }\textbf {\bibinfo
  {volume} {11}},\ \bibinfo {pages} {521} (\bibinfo {year} {1996})}\BibitemShut
  {NoStop}%
\bibitem [{\citenamefont {Blasberg}\ and\ \citenamefont
  {Suter}(1994)}]{Blasberg1994}%
  \BibitemOpen
  \bibfield  {author} {\bibinfo {author} {\bibfnamefont {T.}~\bibnamefont
  {Blasberg}}\ and\ \bibinfo {author} {\bibfnamefont {D.}~\bibnamefont
  {Suter}},\ }\href {\doibase 10.1016/0030-4018(94)90750-1} {\bibfield
  {journal} {\bibinfo  {journal} {Optics Communications}\ }\textbf {\bibinfo
  {volume} {109}},\ \bibinfo {pages} {133} (\bibinfo {year}
  {1994})}\BibitemShut {NoStop}%
\bibitem [{\citenamefont {Blasberg}\ and\ \citenamefont
  {Suter}(1995)}]{Blasberg1995}%
  \BibitemOpen
  \bibfield  {author} {\bibinfo {author} {\bibfnamefont {T.}~\bibnamefont
  {Blasberg}}\ and\ \bibinfo {author} {\bibfnamefont {D.}~\bibnamefont
  {Suter}},\ }\href@noop {} {\bibfield  {journal} {\bibinfo  {journal}
  {Physical Review B}\ }\textbf {\bibinfo {volume} {51}},\ \bibinfo {pages}
  {6309} (\bibinfo {year} {1995})}\BibitemShut {NoStop}%
\bibitem [{\citenamefont {Walther}\ \emph {et~al.}(2016)\citenamefont
  {Walther}, \citenamefont {Nilsson}, \citenamefont {Li}, \citenamefont
  {Rippe},\ and\ \citenamefont {Kr{\"{o}}ll}}]{Walther2016}%
  \BibitemOpen
  \bibfield  {author} {\bibinfo {author} {\bibfnamefont {A.}~\bibnamefont
  {Walther}}, \bibinfo {author} {\bibfnamefont {A.~N.}\ \bibnamefont
  {Nilsson}}, \bibinfo {author} {\bibfnamefont {Q.}~\bibnamefont {Li}},
  \bibinfo {author} {\bibfnamefont {L.}~\bibnamefont {Rippe}}, \ and\ \bibinfo
  {author} {\bibfnamefont {S.}~\bibnamefont {Kr{\"{o}}ll}},\ }\href {\doibase
  10.1140/epjd/e2016-60716-6} {\bibfield  {journal} {\bibinfo  {journal}
  {European Physical Journal D}\ }\textbf {\bibinfo {volume} {70}} (\bibinfo
  {year} {2016})}\BibitemShut {NoStop}%
\bibitem [{\citenamefont {Serrano}\ \emph {et~al.}(2017)\citenamefont
  {Serrano}, \citenamefont {Karlsson}, \citenamefont {Fossati}, \citenamefont
  {Ferrier},\ and\ \citenamefont {Goldner}}]{Serrano2017}%
  \BibitemOpen
  \bibfield  {author} {\bibinfo {author} {\bibfnamefont {D.}~\bibnamefont
  {Serrano}}, \bibinfo {author} {\bibfnamefont {J.}~\bibnamefont {Karlsson}},
  \bibinfo {author} {\bibfnamefont {A.}~\bibnamefont {Fossati}}, \bibinfo
  {author} {\bibfnamefont {A.}~\bibnamefont {Ferrier}}, \ and\ \bibinfo
  {author} {\bibfnamefont {P.}~\bibnamefont {Goldner}},\ }\href
  {http://arxiv.org/abs/1711.03934} {}\Eprint {http://arxiv.org/abs/1711.03934}
  {arXiv:1711.03934}  (\bibinfo {year} {2017})\BibitemShut {NoStop}%
\bibitem [{\citenamefont {Koster}\ \emph {et~al.}(1963)\citenamefont {Koster},
  \citenamefont {Dimmock}, \citenamefont {Wheeler},\ and\ \citenamefont
  {Statz}}]{Koster1963}%
  \BibitemOpen
  \bibfield  {author} {\bibinfo {author} {\bibfnamefont {G.~F.}\ \bibnamefont
  {Koster}}, \bibinfo {author} {\bibfnamefont {J.~O.}\ \bibnamefont {Dimmock}},
  \bibinfo {author} {\bibfnamefont {R.~G.}\ \bibnamefont {Wheeler}}, \ and\
  \bibinfo {author} {\bibfnamefont {H.}~\bibnamefont {Statz}},\ }\href@noop {}
  {\emph {\bibinfo {title} {{Properties of the thirty-two point groups}}}}\
  (\bibinfo  {publisher} {The MIT Press},\ \bibinfo {address} {Cambridge, MA},\
  \bibinfo {year} {1963})\BibitemShut {NoStop}%
\bibitem [{\citenamefont {{Di Bartolo}}(1968)}]{DiBartolo1968}%
  \BibitemOpen
  \bibfield  {author} {\bibinfo {author} {\bibfnamefont {B.}~\bibnamefont {{Di
  Bartolo}}},\ }\href {\doibase 10.1088/1367-2630/18/1/013006} {\emph {\bibinfo
  {title} {{Optical interactions in solids}}}}\ (\bibinfo  {publisher}
  {Wiley},\ \bibinfo {address} {New York},\ \bibinfo {year} {1968})\BibitemShut
  {NoStop}%
\bibitem [{\citenamefont {Henderson}\ and\ \citenamefont
  {Imbush}(2006)}]{Henderson2006}%
  \BibitemOpen
  \bibfield  {author} {\bibinfo {author} {\bibfnamefont {B.}~\bibnamefont
  {Henderson}}\ and\ \bibinfo {author} {\bibfnamefont {G.}~\bibnamefont
  {Imbush}},\ }\href {\doibase 10.1016/S1369-7021(06)71623-6} {\emph {\bibinfo
  {title} {{Optical spectroscopy of inorganic solids}}}}\ (\bibinfo
  {publisher} {Oxford University Press},\ \bibinfo {year} {2006})\BibitemShut
  {NoStop}%
\bibitem [{\citenamefont {Shi}\ \emph {et~al.}(2001)\citenamefont {Shi},
  \citenamefont {Zhang},\ and\ \citenamefont {Shen}}]{Shi2001}%
  \BibitemOpen
  \bibfield  {author} {\bibinfo {author} {\bibfnamefont {H.~S.}\ \bibnamefont
  {Shi}}, \bibinfo {author} {\bibfnamefont {G.}~\bibnamefont {Zhang}}, \ and\
  \bibinfo {author} {\bibfnamefont {H.~Y.}\ \bibnamefont {Shen}},\ }\href@noop
  {} {\bibfield  {journal} {\bibinfo  {journal} {J. Synth. Cryst}\ }\textbf
  {\bibinfo {volume} {30}},\ \bibinfo {pages} {85} (\bibinfo {year}
  {2001})}\BibitemShut {NoStop}%
\bibitem [{\citenamefont {Purcell}(1946)}]{Purcell1946}%
  \BibitemOpen
  \bibfield  {author} {\bibinfo {author} {\bibfnamefont {E.~M.}\ \bibnamefont
  {Purcell}},\ }\href {http://www.citeulike.org/group/2870/article/2154423}
  {\bibfield  {journal} {\bibinfo  {journal} {Physical Review}\ }\textbf
  {\bibinfo {volume} {69}},\ \bibinfo {pages} {681} (\bibinfo {year}
  {1946})}\BibitemShut {NoStop}%
\bibitem [{\citenamefont {Sumida}\ and\ \citenamefont {Fan}(1994)}]{Fan1994}%
  \BibitemOpen
  \bibfield  {author} {\bibinfo {author} {\bibfnamefont {D.~S.}\ \bibnamefont
  {Sumida}}\ and\ \bibinfo {author} {\bibfnamefont {T.~Y.}\ \bibnamefont
  {Fan}},\ }\href {\doibase 10.1364/OL.19.001343} {\bibfield  {journal}
  {\bibinfo  {journal} {Optics Letters}\ }\textbf {\bibinfo {volume} {19}},\
  \bibinfo {pages} {1343} (\bibinfo {year} {1994})}\BibitemShut {NoStop}%
\bibitem [{\citenamefont {Kisel}\ \emph {et~al.}(2004)\citenamefont {Kisel},
  \citenamefont {Troshin}, \citenamefont {Tolstik}, \citenamefont
  {Shcherbitsky}, \citenamefont {Kuleshov}, \citenamefont {Matrosov},
  \citenamefont {Matrosova},\ and\ \citenamefont {Kupchenko}}]{Kisel2004}%
  \BibitemOpen
  \bibfield  {author} {\bibinfo {author} {\bibfnamefont {V.~E.}\ \bibnamefont
  {Kisel}}, \bibinfo {author} {\bibfnamefont {S.~E.}\ \bibnamefont {Troshin}},
  \bibinfo {author} {\bibfnamefont {N.~S.}\ \bibnamefont {Tolstik}}, \bibinfo
  {author} {\bibfnamefont {V.~G.}\ \bibnamefont {Shcherbitsky}}, \bibinfo
  {author} {\bibfnamefont {N.~V.}\ \bibnamefont {Kuleshov}}, \bibinfo {author}
  {\bibfnamefont {V.~N.}\ \bibnamefont {Matrosov}}, \bibinfo {author}
  {\bibfnamefont {T.~S.}\ \bibnamefont {Matrosova}}, \ and\ \bibinfo {author}
  {\bibfnamefont {M.~I.}\ \bibnamefont {Kupchenko}},\ }\href {\doibase
  10.1364/OL.29.002491} {\bibfield  {journal} {\bibinfo  {journal} {Optics
  Letters}\ }\textbf {\bibinfo {volume} {29}},\ \bibinfo {pages} {2491}
  (\bibinfo {year} {2004})}\BibitemShut {NoStop}%
\bibitem [{\citenamefont {B{\"{o}}ttger}\ \emph {et~al.}(2006)\citenamefont
  {B{\"{o}}ttger}, \citenamefont {Thiel}, \citenamefont {Sun},\ and\
  \citenamefont {Cone}}]{Bottger2006}%
  \BibitemOpen
  \bibfield  {author} {\bibinfo {author} {\bibfnamefont {T.}~\bibnamefont
  {B{\"{o}}ttger}}, \bibinfo {author} {\bibfnamefont {C.~W.}\ \bibnamefont
  {Thiel}}, \bibinfo {author} {\bibfnamefont {Y.}~\bibnamefont {Sun}}, \ and\
  \bibinfo {author} {\bibfnamefont {R.~L.}\ \bibnamefont {Cone}},\ }\href
  {\doibase 10.1103/PhysRevB.73.075101} {\bibfield  {journal} {\bibinfo
  {journal} {Physical Review B}\ }\textbf {\bibinfo {volume} {73}},\ \bibinfo
  {pages} {075101} (\bibinfo {year} {2006})}\BibitemShut {NoStop}%
\bibitem [{\citenamefont {Mims}(1968)}]{Mims1968}%
  \BibitemOpen
  \bibfield  {author} {\bibinfo {author} {\bibfnamefont {W.}~\bibnamefont
  {Mims}},\ }\href {\doibase 10.1103/PhysRev.168.370} {\bibfield  {journal}
  {\bibinfo  {journal} {Physical Review}\ }\textbf {\bibinfo {volume} {168}},\
  \bibinfo {pages} {370} (\bibinfo {year} {1968})}\BibitemShut {NoStop}%
\bibitem [{\citenamefont {Sun}\ \emph {et~al.}(2002)\citenamefont {Sun},
  \citenamefont {Thiel}, \citenamefont {Cone}, \citenamefont {Equall},\ and\
  \citenamefont {Hutcheson}}]{Sun2002}%
  \BibitemOpen
  \bibfield  {author} {\bibinfo {author} {\bibfnamefont {Y.}~\bibnamefont
  {Sun}}, \bibinfo {author} {\bibfnamefont {C.~W.}\ \bibnamefont {Thiel}},
  \bibinfo {author} {\bibfnamefont {R.~L.}\ \bibnamefont {Cone}}, \bibinfo
  {author} {\bibfnamefont {R.~W.}\ \bibnamefont {Equall}}, \ and\ \bibinfo
  {author} {\bibfnamefont {R.~L.}\ \bibnamefont {Hutcheson}},\ }\href {\doibase
  10.1016/S0022-2313(02)00281-8} {\bibfield  {journal} {\bibinfo  {journal}
  {Journal of Luminescence}\ }\textbf {\bibinfo {volume} {98}},\ \bibinfo
  {pages} {281} (\bibinfo {year} {2002})}\BibitemShut {NoStop}%
\bibitem [{\citenamefont {Ganem}\ \emph {et~al.}(1991)\citenamefont {Ganem},
  \citenamefont {Wang}, \citenamefont {Boye}, \citenamefont {Meltzer},
  \citenamefont {Yen},\ and\ \citenamefont {Macfarlane}}]{Ganem1991}%
  \BibitemOpen
  \bibfield  {author} {\bibinfo {author} {\bibfnamefont {J.}~\bibnamefont
  {Ganem}}, \bibinfo {author} {\bibfnamefont {Y.~P.}\ \bibnamefont {Wang}},
  \bibinfo {author} {\bibfnamefont {D.}~\bibnamefont {Boye}}, \bibinfo {author}
  {\bibfnamefont {R.~S.}\ \bibnamefont {Meltzer}}, \bibinfo {author}
  {\bibfnamefont {W.~M.}\ \bibnamefont {Yen}}, \ and\ \bibinfo {author}
  {\bibfnamefont {R.~M.}\ \bibnamefont {Macfarlane}},\ }\href {\doibase
  10.1103/PhysRevLett.61.2596} {\bibfield  {journal} {\bibinfo  {journal}
  {Physical Review Letters}\ }\textbf {\bibinfo {volume} {66}},\ \bibinfo
  {pages} {695} (\bibinfo {year} {1991})}\BibitemShut {NoStop}%
\bibitem [{\citenamefont {Rakonjac}\ \emph {et~al.}(2018)\citenamefont
  {Rakonjac}, \citenamefont {Chen}, \citenamefont {Horvath},\ and\
  \citenamefont {Longdell}}]{Rakonjac2018a}%
  \BibitemOpen
  \bibfield  {author} {\bibinfo {author} {\bibfnamefont {J.~V.}\ \bibnamefont
  {Rakonjac}}, \bibinfo {author} {\bibfnamefont {Y.-H.}\ \bibnamefont {Chen}},
  \bibinfo {author} {\bibfnamefont {S.~P.}\ \bibnamefont {Horvath}}, \ and\
  \bibinfo {author} {\bibfnamefont {J.~J.}\ \bibnamefont {Longdell}},\ }\href
  {http://arxiv.org/abs/1802.03862} {}\Eprint {http://arxiv.org/abs/1802.03862}
  {arXiv:1802.03862}  (\bibinfo {year} {2018})\BibitemShut {NoStop}%
\bibitem [{\citenamefont {Fisk}\ \emph {et~al.}(1990)\citenamefont {Fisk},
  \citenamefont {He}, \citenamefont {Holliday},\ and\ \citenamefont
  {Manson}}]{Fisk1990}%
  \BibitemOpen
  \bibfield  {author} {\bibinfo {author} {\bibfnamefont {P.~T.~H.}\
  \bibnamefont {Fisk}}, \bibinfo {author} {\bibfnamefont {X.~F.}\ \bibnamefont
  {He}}, \bibinfo {author} {\bibfnamefont {K.}~\bibnamefont {Holliday}}, \ and\
  \bibinfo {author} {\bibfnamefont {N.~B.}\ \bibnamefont {Manson}},\ }\href
  {\doibase 10.1016/0022-2313(90)90096-T} {\bibfield  {journal} {\bibinfo
  {journal} {Journal of Luminescence}\ }\textbf {\bibinfo {volume} {45}},\
  \bibinfo {pages} {26} (\bibinfo {year} {1990})}\BibitemShut {NoStop}%
\bibitem [{\citenamefont {Karlsson}\ \emph {et~al.}(2017)\citenamefont
  {Karlsson}, \citenamefont {Kunkel}, \citenamefont {Ikesue}, \citenamefont
  {Ferrier},\ and\ \citenamefont {Goldner}}]{Karlsson2017}%
  \BibitemOpen
  \bibfield  {author} {\bibinfo {author} {\bibfnamefont {J.}~\bibnamefont
  {Karlsson}}, \bibinfo {author} {\bibfnamefont {N.}~\bibnamefont {Kunkel}},
  \bibinfo {author} {\bibfnamefont {A.}~\bibnamefont {Ikesue}}, \bibinfo
  {author} {\bibfnamefont {A.}~\bibnamefont {Ferrier}}, \ and\ \bibinfo
  {author} {\bibfnamefont {P.}~\bibnamefont {Goldner}},\ }\href {\doibase
  10.1088/1361-648X/aa529a} {\bibfield  {journal} {\bibinfo  {journal} {Journal
  of Physics Condensed Matter}\ }\textbf {\bibinfo {volume} {29}} (\bibinfo
  {year} {2017})}\BibitemShut {NoStop}%
\bibitem [{\citenamefont {Ahlefeldt}\ \emph {et~al.}(2016)\citenamefont
  {Ahlefeldt}, \citenamefont {Hush},\ and\ \citenamefont
  {Sellars}}]{Ahlefeldt2016}%
  \BibitemOpen
  \bibfield  {author} {\bibinfo {author} {\bibfnamefont {R.~L.}\ \bibnamefont
  {Ahlefeldt}}, \bibinfo {author} {\bibfnamefont {M.~R.}\ \bibnamefont {Hush}},
  \ and\ \bibinfo {author} {\bibfnamefont {M.~J.}\ \bibnamefont {Sellars}},\
  }\href {\doibase 10.1103/PhysRevLett.117.250504} {\bibfield  {journal}
  {\bibinfo  {journal} {Physical Review Letters}\ }\textbf {\bibinfo {volume}
  {117}},\ \bibinfo {pages} {250504} (\bibinfo {year} {2016})}\BibitemShut
  {NoStop}%
\bibitem [{\citenamefont {Zhong}\ \emph
  {et~al.}(2015{\natexlab{b}})\citenamefont {Zhong}, \citenamefont {Kindem},
  \citenamefont {Miyazono},\ and\ \citenamefont {Faraon}}]{TZhong2015}%
  \BibitemOpen
  \bibfield  {author} {\bibinfo {author} {\bibfnamefont {T.}~\bibnamefont
  {Zhong}}, \bibinfo {author} {\bibfnamefont {J.~M.}\ \bibnamefont {Kindem}},
  \bibinfo {author} {\bibfnamefont {E.}~\bibnamefont {Miyazono}}, \ and\
  \bibinfo {author} {\bibfnamefont {A.}~\bibnamefont {Faraon}},\ }\href
  {\doibase 10.1038/ncomms9206} {\bibfield  {journal} {\bibinfo  {journal}
  {Nature Communications}\ }\textbf {\bibinfo {volume} {6}},\ \bibinfo {pages}
  {8206} (\bibinfo {year} {2015}{\natexlab{b}})}\BibitemShut {NoStop}%
\bibitem [{\citenamefont {Bartholomew}\ \emph {et~al.}(2016)\citenamefont
  {Bartholomew}, \citenamefont {Ahlefeldt},\ and\ \citenamefont
  {Sellars}}]{Bartholomew2016}%
  \BibitemOpen
  \bibfield  {author} {\bibinfo {author} {\bibfnamefont {J.~G.}\ \bibnamefont
  {Bartholomew}}, \bibinfo {author} {\bibfnamefont {R.~L.}\ \bibnamefont
  {Ahlefeldt}}, \ and\ \bibinfo {author} {\bibfnamefont {M.~J.}\ \bibnamefont
  {Sellars}},\ }\href {\doibase 10.1103/PhysRevB.93.014401} {\bibfield
  {journal} {\bibinfo  {journal} {Physical Review B}\ }\textbf {\bibinfo
  {volume} {93}},\ \bibinfo {pages} {014401} (\bibinfo {year}
  {2016})}\BibitemShut {NoStop}%
\bibitem [{\citenamefont {Elliott}\ and\ \citenamefont
  {Stevens}(1953)}]{Elliott1953}%
  \BibitemOpen
  \bibfield  {author} {\bibinfo {author} {\bibfnamefont {R.~J.}\ \bibnamefont
  {Elliott}}\ and\ \bibinfo {author} {\bibfnamefont {K.~W.~H.}\ \bibnamefont
  {Stevens}},\ }\href {\doibase 10.1098/rspa.1953.0124} {\bibfield  {journal}
  {\bibinfo  {journal} {Proceedings of the Royal Society A: Mathematical,
  Physical and Engineering Sciences}\ }\textbf {\bibinfo {volume} {218}},\
  \bibinfo {pages} {553} (\bibinfo {year} {1953})}\BibitemShut {NoStop}%
\end{thebibliography}%

\end{document}